\documentclass[12pt]{article}
\usepackage{amsfonts}
\usepackage{graphicx}

\begin{document}

\title{Wavefronts and Light Cones for Kerr Spacetimes}

\author{Francisco Frutos-Alfaro\footnote{Space Research Center, 
University of Costa Rica, San Jos\'e, Costa Rica, 
email: frutos@fisica.ucr.ac.cr} \\
Frank Grave\footnote{Institute for Visualization und Interactive Systems, 
University of Stuttgart, Universit\"atsstrasse 38, 70569 Stuttgart, Germany} \\
Thomas M\"uller\footnote{Institute for Visualization und Interactive Systems, 
University of Stuttgart, Universit\"atsstrasse 38, 70569 Stuttgart, Germany} \\ 
Daria Adis\footnote{Theoretical Astrophysics, University of T\"ubingen, 
Auf der Morgenstelle 10C, 72076 T\"ubingen, Germany}}
 
\date{\today}

\maketitle

\begin{abstract}
We investigate the light propagation by means of simulations of wavefronts and 
light cones for Kerr spacetimes. Simulations of this kind give us a new insight 
to better understand the light propagation in presence of massive rotating 
black holes. A relevant result is that wavefronts are back scattered with 
winding around the black hole. To generate these visualizations, an 
interactive computer program with a graphical user interface, called 
{\texttt{JWFront}}, was written in {\texttt{Java}}.
\end{abstract}

\bigskip

\section{Introduction}

\noindent
In general relativity and astrophysics, the Kerr metric is useful to study, 
for example, stellar compact objects, like accreation disks in neutron stars.
This metric was found by Kerr in 1963 [1], since then this spacetime appears 
in many articles on these topics and it is currently one of the most used 
spacetimes, because it represents a spacetime of a massive rotating object.

\noindent
Friedrich and Stewart [2, 3] based on Arnold's catastrophe theory [4] 
developed the theory of wavefronts and singularities (caustics) in 
general relativity. Recently, Hasse [5] {\it et al}., Low [6], and Ehlers and 
Newman [7] have revived this topic from a mathematical viewpoint. The wavefront 
propagation, caustic and the light cone structures for a non rotating object, 
described with the Schwarzschild spacetime, was discussed by Perlick [8]. 
Caustics for the Kerr metric were numerically computed by Rauch and Blandford 
[9]. Grave studied the gravitational collapse and wavefronts for this 
spacetime [10]. More recently, Sereno and De Luca [11] computed these caustics 
using a Taylor expansion of lightlike geodesics. A numerical treatment on the 
structure of Kerr caustics was done by Bozza [12]. Qualitative descriptions of 
wavefronts and caustics for gravitational lensing were presented by Blandford 
and Narayan [13], Schneider {\it et al}. [14], and Ohanian and Ruffini [15]. 
Petters {\it et al}. [16, 17], and Frittelli and Petters [18] addressed 
formally this subject. Ellis {\it et al}. [19] discussed qualitatively the 
light cone structure for gravitational lensing.

\noindent
In this work, wavefronts, caustics and light cones for the Kerr spacetime are 
investigated. Nowadays, computer simulations are becoming relevant in 
general relativity, because they can help understand complex phenomena. 
With the new technologies, these simulations can practically be done in real 
time. Thus, the aim of this work is to provide a new perspective about 
wavefront propagations in the Kerr spacetime by means of computer simulation. 
For this purpose, we have designed {\texttt{JWFront}} [20], an interactive 
{\texttt{Java}} program using {\texttt{OpenGL}} (Open Graphics Library), 
to visualize wavefronts and light cones for these spacetimes.   

\noindent
In the next section, the Kerr spacetime and its tetrads will be briefly 
introduced and discussed. The equation of motion, i.e. the geodesic 
equation and how it is solved, will be showed in the third section. 
Definitions for the sake of visualizations as well as discussions about the 
wavefront, caustic and lightcone structures are presented in the fourth 
section. A succinct discussion about our program {\texttt{JWFront}} will be 
given in the fifth section. The last section is devoted to discuss the results 
of the visualizations for the Kerr spacetimes.

\section{Kerr Spacetimes}

\noindent
The Kerr metric is an exact solution of the vacuum Einstein field equation and 
represents the spacetime of a massive rotating black hole. In this spacetime, 
the rotating body would exhibit an {\it inertial frame dragging} 
(Lense-Thirring effect), i.e., a particle moving close to it would corotate. 
This is not because of any force or torque applied on the particle, but rather 
because of the spacetime curvature associated with this black hole. 
This region is called the {\it ergosphere}. At large distances this spacetime 
is flat (asymptotically flat). In Boyer-Linquist coordinates the metric 
has the following form [21, 22]:

\begin{eqnarray}
\label{metric}
d s^2 & = & - \frac{\Delta}{\rho^2} [d t - a \sin^2{\theta} d \phi]^2 \\ 
& + &\frac{\sin^2{\theta}}{\rho^2} [(r^2 + a^2) d \phi - a d t]^2 
+ \frac{\rho^2}{\Delta} d r^2 + \rho^2 d \theta^2  \nonumber \\
& = & g_{0 0} d t^2 + 2 g_{0 3} dt d \phi + g_{1 1} d r^2 
+ g_{2 2} d \theta^2 + g_{3 3} d \phi^2 , \nonumber
\end{eqnarray}

\noindent
where $ \Delta = r^2 - R_s r + a^2 $, $ \rho^2 = r^2 + a^2 \cos^2{\theta} $, 
$ R_s = 2 M $ is the Schwarzschild radius in geometrical units 
($ c = G = 1$), $ M $ is the mass of the black hole, $ a = J / M $ 
(angular momentum per unit mass, $ a \le M $), and $ g_{\mu \nu} $ are 
the metric components, which can be read easily from (\ref{metric}). 
The Kerr spacetime contains the Minkowski flat metric, the Schwarzschild 
metric, and the Lense-Thirring spacetime. If $ a^2 \sim 0 $, i.e. neglecting 
the second order in powers of $ a $, one gets the Lense-Thirring metric, which 
represents the metric of a massive slow-rotating body. We get the Schwarzschild 
metric if $ a = 0 $, which represents the metric of a massive non-rotating body.


\bigskip
\noindent
{\bf Local Frames: Tetrad Formalism}
\smallskip

\noindent
The tetrad formalism is very useful in general relativity. It defines 
a mathematical element called tetrad or {\it vierbein}, which is used to 
connect the curved coordinate systems with the local flat Lorentz coordinates. 
These tetrads must fulfill the equation

\begin{equation}
\label{tetrad}
\eta_{(\alpha) (\beta)} = g_{\mu \nu} e^{\mu}_{(\alpha)} e^{\nu}_{(\beta)} ,
\end{equation}

\noindent
where $ e^{\mu}_{(\beta)} $ is a chosen vierbein element, 
$ \eta_{(\alpha) (\beta)} $ stands for the Min\-kows\-ki me\-tric 
$ ({\rm diag} (- 1, \, 1, \, 1, \, 1)) $.

\noindent
For the Kerr spacetime, there are at least two possibilities to choose 
the tetrads. The first one is called the locally static frame (LSF). 
In this frame, the observer is static. This kind of observer cannot be located 
in the ergosphere, because they would move with superluminal velocity in this 
region to counteract the Lense-Thirring effect. The local tetrads for this 
static observer have following components:

\begin{eqnarray}
\label{lsf}
e_{(0)} & = & \frac{1}{\sqrt{- g_{0 0}}} \frac{\partial}{\partial t} 
= \frac{\rho}{\sqrt{\rho^2 - R_s r}} \frac{\partial}{\partial t} \nonumber \\
e_{(1)} & = & \frac{1}{\sqrt{g_{1 1}}} \frac{\partial}{\partial r} 
= \frac{\sqrt{\Delta}}{\rho} \frac{\partial}{\partial r} \\
e_{(2)} & = & \frac{1}{\sqrt{g_{2 2}}} \frac{\partial}{\partial \theta} 
= \frac{1}{\rho} \frac{\partial}{\partial \theta} \nonumber \\
e_{(3)} & = & - \frac{g_{0 3}}{\sqrt{g_{0 0} 
(g_{0 0} g_{3 3} - g^2_{0 3})}} \frac{\partial}{\partial t}  
+ \sqrt{\frac{g_{0 0}}{g_{0 0} g_{3 3} 
- g^2_{0 3}}} \frac{\partial}{\partial \phi}  \nonumber \\
& = & \frac{R_s a r \sin{\theta}}{\rho \sqrt{\Delta (\rho^2 - R_s r)}} 
\frac{\partial}{\partial t} + \frac{1}{\rho \sin{\theta}} 
\sqrt{\frac{\rho^2 - R_s r}{\Delta}} \frac{\partial}{\partial \phi} , 
\nonumber 
\end{eqnarray}

\noindent
where $ {\partial} / {\partial t}, \, {\partial} / {\partial r}, 
\, {\partial} / {\partial \theta} $ and $ {\partial} / {\partial \phi} $ are 
understood as unit vector directions.

\noindent
The second one is called locally nonrotating frame [23] (LNRF), in which 
the observer is stationary. An observer in this kind of frames 
could be in the ergosphere. The local tetrads for this stationary observer 
have following components:

\begin{eqnarray}
\label{lnrf}
e_{(0)} & = & \sqrt{\frac{g_{3 3}}{{g^2_{0 3} 
- g_{0 0} g_{3 3}}}} \frac{\partial}{\partial t} 
- \frac{g_{0 3}}{\sqrt{g_{3 3}(g^2_{0 3} 
- g_{0 0} g_{3 3})}} \frac{\partial}{\partial \phi} \nonumber \\
& = & \frac{\Sigma}{\rho \sqrt{\Delta}} \frac{\partial}{\partial t} 
+ \frac{R_s a r}{\rho \Sigma \sqrt{\Delta}} 
\frac{\partial}{\partial \phi} \nonumber \\
e_{(1)} & = & \frac{1}{\sqrt{g_{1 1}}} \frac{\partial}{\partial r} 
= \frac{\sqrt{\Delta}}{\rho} \frac{\partial}{\partial r} \\
e_{(2)} & = & \frac{1}{\sqrt{g_{2 2}}} \frac{\partial}{\partial \theta} 
= \frac{1}{\rho} \frac{\partial}{\partial \theta} \nonumber \\
e_{(3)} & = & \frac{1}{\sqrt{{g_{3 3}}}} \frac{\partial}{\partial \phi} 
= \frac{\rho}{\Sigma \sin{\theta}} 
\frac{\partial}{\partial \phi} , \nonumber \\
\end{eqnarray}

\noindent
where $ \Sigma^2 = (r^2 + a^2)^2 - a^2 \Delta \sin^2{\theta} $. 
These tetrad definitions are useful to find the trajectories of light rays 
moving in a Kerr spacetime.

\section{The Geodesic Equation and Its Solution}

\noindent
In general relativity, the trajectory of particles or light rays can be 
determined by the geodesic equation. Generally, this equation can only be 
solved using numerical methods. 
This equation has the following form [21, 22]:

\begin{equation}
\label{geo}
\frac{d^2 x^{\mu}}{d \lambda^2} + \Gamma^{\mu}_{\alpha \beta} 
\frac{d x^{\alpha}}{d \lambda} \frac{d x^{\beta}}{d \lambda} = 0 ,
\end{equation}

\noindent
where $ \alpha, \, \beta, \, \mu = 0, \, 1, \, 2, \, 3 $, and $ \lambda $ 
is an affine parameter, a parameter such that $ {d x^{\beta}}/{d \lambda} $ 
has constant magnitude (affine parametrization). The components 
$ \Gamma^{\mu}_{\alpha \beta} $, called the Christoffel symbols, are given by 
 
\begin{equation}
\label{christofsymbol}
\Gamma^{\mu}_{\alpha \beta} = 
\frac{g^{\mu \nu}}{2} \left\{
\frac{\partial g_{\alpha \nu}}{\partial x^{\beta}} 
+ \frac{\partial g_{\beta \nu}}{\partial x^{\alpha}} 
- \frac{\partial g_{\alpha \beta}}{\partial x^{\nu}} \right\} .
\end{equation}

\noindent
These symbols for the Kerr metric can be computed by means of symbolic 
programs. A program using the free symbolic software {\sf Reduce} [24] was 
written to obtain them. In the Appendix, the non-null Christoffel symbols are 
listed. Introducing these Christoffel symbols into equation (\ref{geo}), 
we have four ordinary second order differential equations given by:

\begin{eqnarray}
\label{geo2}
\frac{d^2 t}{d \lambda^2} 
+ 2 \frac{d t}{d \lambda} \left(\Gamma^{0}_{0 1} \frac{d r}{d \lambda} 
+ \Gamma^{0}_{0 2} \frac{d \theta}{d \lambda} \right)
+ 2 \frac{d \phi}{d \lambda} \left(\Gamma^{0}_{1 3} \frac{d r}{d \lambda} 
+ \Gamma^{0}_{2 3} \frac{d \theta}{d \lambda} \right)
& = & 0 , \nonumber \\
\frac{d^2 r}{d \lambda^2} 
+ \Gamma^{1}_{0 0} \left(\frac{d t}{d \lambda} \right)^2 
+ 2 \Gamma^{1}_{0 3} \frac{d t}{d \lambda} \frac{d \phi}{d \lambda} 
+ \Gamma^{1}_{1 1} \left(\frac{d r}{d \lambda} \right)^2  \nonumber \\
+ 2 \Gamma^{1}_{1 2} \frac{d r}{d \lambda} \frac{d \theta}{d \lambda} 
+ \Gamma^{1}_{2 2} \left(\frac{d \theta}{d \lambda} \right)^2
+ \Gamma^{1}_{3 3} \left(\frac{d \phi}{d \lambda} \right)^2 
& = & 0 , \\
\frac{d^2 \theta}{d \lambda^2} 
+ \Gamma^{2}_{0 0} \left(\frac{d t}{d \lambda} \right)^2 
+ 2 \Gamma^{2}_{0 3} \frac{d t}{d \lambda} \frac{d \phi}{d \lambda} 
+ \Gamma^{2}_{1 1} \left(\frac{d r}{d \lambda} \right)^2 \nonumber \\
+ 2 \Gamma^{2}_{1 2} \frac{d r}{d \lambda} \frac{d \theta}{d \lambda} 
+ \Gamma^{2}_{2 2} \left(\frac{d \theta}{d \lambda} \right)^2 
+ \Gamma^{2}_{3 3} \left(\frac{d \phi}{d \lambda} \right)^2  
& = & 0 , \nonumber \\
\frac{d^2 \phi}{d \lambda^2} 
+ 2 \frac{d t}{d \lambda} \left(\Gamma^{3}_{0 1} \frac{d r}{d \lambda} 
+ \Gamma^{3}_{0 2} \frac{d \theta}{d \lambda} \right)
+ 2 \frac{d \phi}{d \lambda} \left(\Gamma^{3}_{1 3} \frac{d r}{d \lambda} 
+ \Gamma^{3}_{2 3} \frac{d \theta}{d \lambda} \right) 
& = & 0 . \nonumber 
\end{eqnarray}

\noindent
For light rays, there is also another equation they have to 
fulfill, the null geodesic equation (lightlike geodesics):

\begin{eqnarray}
\label{nulgeodesic}
\left(\frac{d s}{d \lambda} \right)^2 
& = & g_{\mu \nu} \frac{d x^{\mu}}{d \lambda} \frac{d x^{\nu}}{d \lambda} 
\nonumber \\
& = & g_{0 0} \left(\frac{d t}{d \lambda} \right)^2 
+ 2 g_{0 3} \left(\frac{d t}{d \lambda} \right) 
\left(\frac{d \phi}{d \lambda} \right) \\
& + & g_{1 1} \left(\frac{d r}{d \lambda} \right)^2 
+ g_{2 2} \left(\frac{d \theta}{d \lambda} \right)^2 
+ g_{3 3} \left(\frac{d \phi}{d \lambda} \right)^2 = 0 . \nonumber
\end{eqnarray}

\noindent
The four-dimensional trajectories of light rays can be found by solving the 
equations (\ref{geo2}) with the constraint equation (\ref{nulgeodesic}). 
Now, one needs initial conditions in order to solve these equations numerically.


\bigskip
\noindent
{\bf Initial Conditions:}
\smallskip

\noindent
The initial spacetime event $ x^{\mu}_0 = (t_0, \, x_0, \, y_0, \, z_0) $ 
for all geodesics of the bundle defining the wavefront (see below) has to be 
given in order to solve numerically equations (\ref{geo2}) with the constraint 
equation (\ref{nulgeodesic}). For each geodesic of the bundle, 
the four-velocity at the initial point, 

$$ \left[\frac{d x^{\mu}}{d \lambda} \right]_0 = 
\left. \left(\frac{d t}{d \lambda}, \, \frac{d x}{d \lambda}, 
\, \frac{d y}{d \lambda}, \, \frac{d z}{d \lambda} \right) \right|_0 , $$ 

\noindent
determines the direction for each geodesic and 
they are calculated as follows: The tridimensional (3D) initial vector, 

$$ \left[\frac{d \tilde {\bf x}}{d \lambda} \right]_0 = 
\left. \left(\frac{d {\tilde x}}{d \lambda}, \, 
\frac{d {\tilde y}}{d \lambda}, \, 
\frac{d {\tilde z }}{d \lambda} \right) \right|_0 , $$ 

\noindent
for each geodesic in local flat spacetime is given input. Using the null 
geodesic condition for this local metric, the initial time derivative 
$ [d {\tilde t} / d \lambda]_0 $ is determined. Now, we have all components 
in local flat spacetime $ [d {\tilde x}^{\mu} / d \lambda]_0 $. Finally, 
the four-velocity in non-flat spacetime is determined by transforming from 
the local flat spacetime to the non-flat spacetime using the tetrads 
$ e^{\mu}_{\nu} $ for the Kerr metric: 

$$ \left[\frac{d x^{\mu}}{d \lambda} \right]_0 = e^{\mu}_{\nu} 
\left[\frac{d {\tilde x}^{\nu}}{d \lambda} \right]_0 . $$ 

\noindent
With the purpose for simulating wavefronts and light cones in mind, one has to 
choose between the two kinds of observers (see second section). 

\noindent
Now, we have all elements to numerically solve the four ordinary 
equations with these initial conditions. For this goal, a fourth order 
Runge-Kutta procedure is used.

\section{Wavefronts, Caustics and Light Cones}

\noindent
{\bf Wavefronts}
\smallskip

\noindent
Formally speaking, the wavefronts are defined as follows: 
A wavefront is generated by a bundle of light rays orthogonal to a 
spacelike 2-surface in a four-dimensional Lorentzian manifold [5]. 

\noindent
To simulate it, the wavefront is defined as the surface $ {\cal A} $ generated 
by all points of the null geodesic bundle at a given time $ t_i $:

\begin{eqnarray}
{\cal A}(t_i) & = & \left \{ \gamma(t_i) \; | \; \gamma(t_i) \; 
{\rm is \; a \; null \; geodesic \; with} \; \right. \\
& & \left. 
\gamma(t_0) = (t_0, \, x_0, \, y_0, \, z_0), \; t_i \ge t_0 \right \} . 
\nonumber 
\end{eqnarray}

\noindent
Qualitatively speaking, the wavefronts that spread out in all directions 
from the source are spherical at the very beginning and if they are 
approaching a deflector, they get distorted and their sheets develop generally 
singularities, cusp ridges and self intersections or caustics.

\noindent
In gravitational lens theory, it is considered that light ray deflection 
occurs only at the place where the deflector is located 
(thin lens approximation). This approximation is very useful in many 
calculations, specially, if we are dealing with strong lensing. 
Under this consideration, wavefronts propagate spherically without any 
perturbation until the deflector, then wavefronts are distorted by the 
deflector. The general case is completely different, because wavefronts get 
already perturbed before they approach the deflector and can wind around 
the black hole (see Figure 1). An observer which is behind the deflector will 
see different sheets of the same wavefront coming from different directions. 
Then, the observer will think that there are multiple images of the same 
source.

\bigskip
\noindent
{\bf Caustics}
\smallskip

\noindent
A caustic of a wavefront is formally defined as the set of all points 
where the wavefront fails to be an (immersed) submanifold [5].

\noindent
Roughly speaking, a caustic is the envelope of reflected or refracted light 
rays by a curved surface or object. A caustic can be a point, a line or 
a surface. For instance, for the Schwarzschild black hole the caustic is a line 
along of the line of sight, and for the point mass lens or non-rotating 
black hole the caustic is a point in the line of sight. Interesting caustic 
shapes can be found in gravitational lens theory, for example, for some 
elliptical lens models, it is common to find diamond shape caustics. 
Another important point to mention about caustics is that if an observer would 
be on a caustic, he would detect a high light intensity 
(mathematically speaking, it would be infinity). 


\bigskip
\noindent
{\bf Light Cones}
\smallskip

\noindent
The light cone is defined as the surface generated by all points 
$ (t, \, x, \, y, \, z) $, that fulfill the geodesic equation with 
the null geodesic condition for a fixed starting event 
$ (t_0, \, x_0, \, y_0, \, z_0) $. 
To visualize the light cones, one has to suppress one space dimension, using 
for instance the coordinates $ (t, \, x, \, y) $, $ (t, \, x, \, z) $ or 
$ (t, \, y, \, z) $. Light cones can also be used to visualize caustic 
structures [13], because time slices or cuts in the light cones represent 
the development of the wavefront. 
The same differences that appeared in the structures of general and lens 
wavefronts are also expected in light cones. 

\noindent
In the present work, we will mainly concentrate on visualizations of 
wavefronts and light cones. For more mathematical details about wavefronts, 
caustics and light cones, the interested reader may consult the references 
at the end. Details of the simulations will be shown in the sixth section.

\section{{\texttt{JWFront}}}

\noindent
An interactive frontend or GUI (graphical user interface) to visualize 
wavefronts and light cones in general relativity, called {\texttt{JWFront}}, 
was written in {\texttt{Java}} [20]. Basically, on this GUI, the user have 
to enter the initial position values and choose the values for mass and 
angular momentum per unit mass ($ M $ and $ a $). Later, the user can choose 
what to see. Among the applications, the user can get from our program, are:

\begin{itemize}
\item wavefront animations in 2D and 3D,  
\item light cone visualizations.
\end{itemize}

\noindent
The light cones are visualized using the coordinate systems 
$ (t, \, x, \, y) $ or \\ $ (t, \, z, \, x) $. 
All data obtained from solving the equations is processed in our program by 
means of {\texttt{Java}} and {\texttt{OpenGL}} subroutines in order to simulate 
wavefronts and light cones. 

\noindent
Moreover, this {\texttt{Java}} program can be easily modified to simulate \
wavefronts and light cones for other spacetime. The user just has to provide 
the Christoffel symbols into the program.

\noindent
The interested reader may send us a message requesting for the program or 
for more information about it.

\section{Simulations with {\texttt{JWFront}}}

\noindent
Now, let us discuss some examples of the simulations obtained by 
{\texttt{JWFront}} (see Figures 2-7). Figures 2, 4 and 6 are visualizations 
for the Schwarzschild metric with $ M = 1 $. Figures 3, 5, and 7 are 
simulations for the Kerr spacetime with $ M = 1 $ and $ a = 0.9 $.

\noindent
In Figure 2, the 2D simulation of a wavefront (light pulse) moving from 
right to left for the Schwarzschild metric is shown. 
The Schwarzschild radius $ R_s = 2 $ is depicted with a yellow circle. 
This wavefront does not cross the horizon, because the time diverges for an 
incoming photon. In the third frame of Figure 2, an observer located 
at the left intersection point can see the light pulse, for an initial value 
of $ x = 6 $, at two different angles. 
Due to the rotational symmetry of the Schwarzschild spacetime, 
the complete picture the observer gets is an Einstein ring. 
Furthermore, because the wavefront infinitely winds around the black hole 
from left to right and right to left, the observer will not see a continuously 
visible Einstein ring. Instead, he would see a blinking Einstein ring. 
Every {\it blink} marks the time when the wavefront is passing the observer. 
The intensity of these events decreases as the area of the wavefront is 
expanding infinitely. In fact, every observer not located on the $ x $ axis 
in this spacetime would see the initial light pulse as a blinking phenomena, 
but not as an Einstein ring. The distorted wavefronts cross themselves in 
different points on the $ - x $ axis$^{20}$ after passing the deflector. 
These points define the caustic line on this axis (see the third frame of 
Fig. 2). In contrary to lens theory: for the point mass lens, the 
caustic is a point on this axis. 

\noindent
The 2D visualization of a wavefront (light pulse) moving from 
right to left for the Kerr metric is shown in Figure 3. In these frames, 
the inner horizon is displayed as a small filled circle, the ergo-region as 
a bigger circle. Because of the rapidly rotation of the black hole, 
the wavefront is not symmetric in this plane. The black hole rotates 
counter-clockwise, and so that the upper part of the wavefront reaches the 
$ y $ axis earlier than the lower part. Again, an observer located in the 
intersection point of the wavefront with itself (third frame of Fig. 3) can 
see the initial light pulse coming from two direction in this plane. 
In this Figure (see second and third frames), we can also see caustic points, 
but because of the rotation they are not located on the $ - x $ axis.

\noindent
In Figures 4 and 5, the 3D visualizations of a wavefront for the 
Schwarz\-sch\-ild and Kerr metrics are shown. In these Figures, 
the wavefront consists of $ 1 / 8 $ of a sphere defined by the initial local 
directions. Every frame of Figures is included, similarly, in the next one. 
We can see that the wavefront, starting from below the $ z $ 
axis, reaches positive $ z $ values, because of the above {\it winding} 
effect. So that, as explained with the last figures, 
every point of the spacetime (excluding those inside of the black hole) is 
reached by this wavefront. From the fourth frames of Figures 4 and 5, it can 
respectively be seen that the crossover line projection is on the $ - x $ axis 
for the Schwarzschild metric, whilst for the Kerr one, it runs on a line 
inclined with respect to the $ - x $ axis. The caustic points can also seen 
in both Figures.

\noindent
In the next two figures, we have the light cone structure of the 
Schwarz\-sch\-ild and Kerr spacetimes. The Schwarzschild light cone is shown 
in Figure 6 ($ (x, \, y, \, t) $ coordinates). The structures observed in 
Figure 2 can also be seen. Here, we do not project, but use the time data of 
the wavefront to construct the light cone. In Figures 7 
($ (x, \, y, \, t) $ coordinates), the visualizations of the light cones for 
the Kerr metric are shown. The structures observed in these Figures 
(including the caustic points) are similar to the corresponding structures of 
Figures 3. In Figures 6 and 7, we can see the crossover line for the 
Schwarzschild and Kerr metric.

\section{Conclusions}

\noindent
The simulations produced by {\texttt{JWFront}} helps understand the light 
propagation in strong gravitational fields with rotation, such as in Kerr 
spacetimes. An interesting feature of wavefronts propagation appeared: 
the wavefronts are scattered back and wind around the black hole. 
Thus, an observer on the line of sight with the deflector and the source 
would see multiple images, and if the black hole does not rotate, the observer 
would see at least one Einstein ring. For Schwarzschild metric this winding 
effect is symmetric whereas for the Kerr one it is not. {\texttt{JWFront}} 
can also displayed the visualizations of light cones in these spacetimes. 
The results of the wavefront visualizations showed that the same structures 
can also be seen with light cone simulations as expected.

\bigskip
\bigskip
\noindent
{\bf Acknowledgments}
\smallskip

\noindent
F. Frutos-Alfaro would like to thank Dr.~rer.~nat.~Antonio~Banichevich and 
PhD. Herberth Morales for fruitful discussions.

\newpage

\bigskip
\bigskip
\noindent
{\bf Appendix: Christoffel Symbols}
\smallskip

\noindent
The non-zero Christoffel symbols for the Kerr metric are given by 

\begin{eqnarray}
\label{christoffel}
\Gamma^0_{0 1} & = & \frac{R_s}{2 \rho^4 \Delta} (r^2 + a^2) 
(2 r^2 - \rho^2) \nonumber \\ 
\Gamma^0_{0 2} & = & - \frac{2 a J r}{\rho^4} \sin{\theta} \cos{\theta} 
\nonumber \\ 
\Gamma^0_{1 3} & = & - \frac{J \sin^2{\theta}}{\rho^4 \Delta}
[\rho^2 (r^2 - a^2) + 2 r^2 (r^2 + a^2)] \nonumber \\ 
\Gamma^0_{2 3} & = & \frac{2 a^2 J r}{\rho^4} \cos{\theta} \sin^3{\theta} 
\nonumber \\ 
\Gamma^1_{0 0} & = & \frac{R_s \Delta}{2 \rho^6} [2 r^2 - \rho^2] 
\nonumber \\ 
\Gamma^1_{0 3} & = & - \frac{J \Delta}{\rho^6} (2 r^2 - \rho^2) 
\sin^2{\theta} \nonumber \\ 
\Gamma^1_{1 1} & = & \frac{1}{\rho^2 \Delta}  
\left[\rho^2 \left(\frac{R_s}{2} - r \right) + r \Delta \right] 
\nonumber \\ 
\Gamma^1_{1 2} & = & - \frac{a^2}{\rho^2} \sin{\theta} \cos{\theta} 
\nonumber \\ 
\Gamma^1_{2 2} & = & - \frac{r \Delta}{\rho^2} \nonumber \\ 
\Gamma^1_{3 3} & = & - \frac{\Delta \sin^2{\theta}}{\rho^6} 
[r \rho^4 - a J (2 r^2 - \rho^2) \sin^2{\theta}] \nonumber \\ 
\Gamma^2_{0 0} & = & - \frac{2 a J r}{\rho^6} \sin{\theta} \cos{\theta}
\nonumber \\ 
\Gamma^2_{0 3} & = & \frac{2 J r}{\rho^6} (r^2 + a^2) 
\sin{\theta} \cos{\theta} \\ 
\Gamma^2_{1 1} & = & \frac{a^2}{\rho^2 \Delta} \sin{\theta} \cos{\theta} 
\nonumber \\ 
\Gamma^2_{1 2} & = & \frac{r}{\rho^2} \nonumber \\ 
\Gamma^2_{2 2} & = & \Gamma^1_{1 2} \nonumber \\ 
\Gamma^2_{3 3} & = & - \frac{\sin{\theta} \cos{\theta}}{\rho^6} 
[\rho^4 \Delta + R_s r (r^2 + a^2)^2] \nonumber \\ 
\Gamma^3_{0 1} & = & \frac{J}{\rho^4 \Delta} [2 r^2 - \rho^2] 
\nonumber \\ 
\Gamma^3_{0 2} & = & - \frac{2 J r \cos{\theta}}{\rho^4 \sin{\theta}} 
\nonumber \\ 
\Gamma^3_{1 3} & = & \frac{1}{\rho^4 \Delta}[r \rho^2 (\rho^2 - R_s r) 
- a J \sin^2{\theta} (2 r^2 - \rho^2)] \nonumber \\ 
\Gamma^3_{2 3} & = & \frac{\cos{\theta}}{\rho^4 \sin{\theta}}
[\rho^4 + 2 a J r \sin^2{\theta}] \nonumber 
\end{eqnarray}

\noindent
These Christoffel symbols coincide with the ones obtained by Semer\'ak$^{24}$.


\begin{figure*}
\centering{
{\includegraphics[scale=0.5]{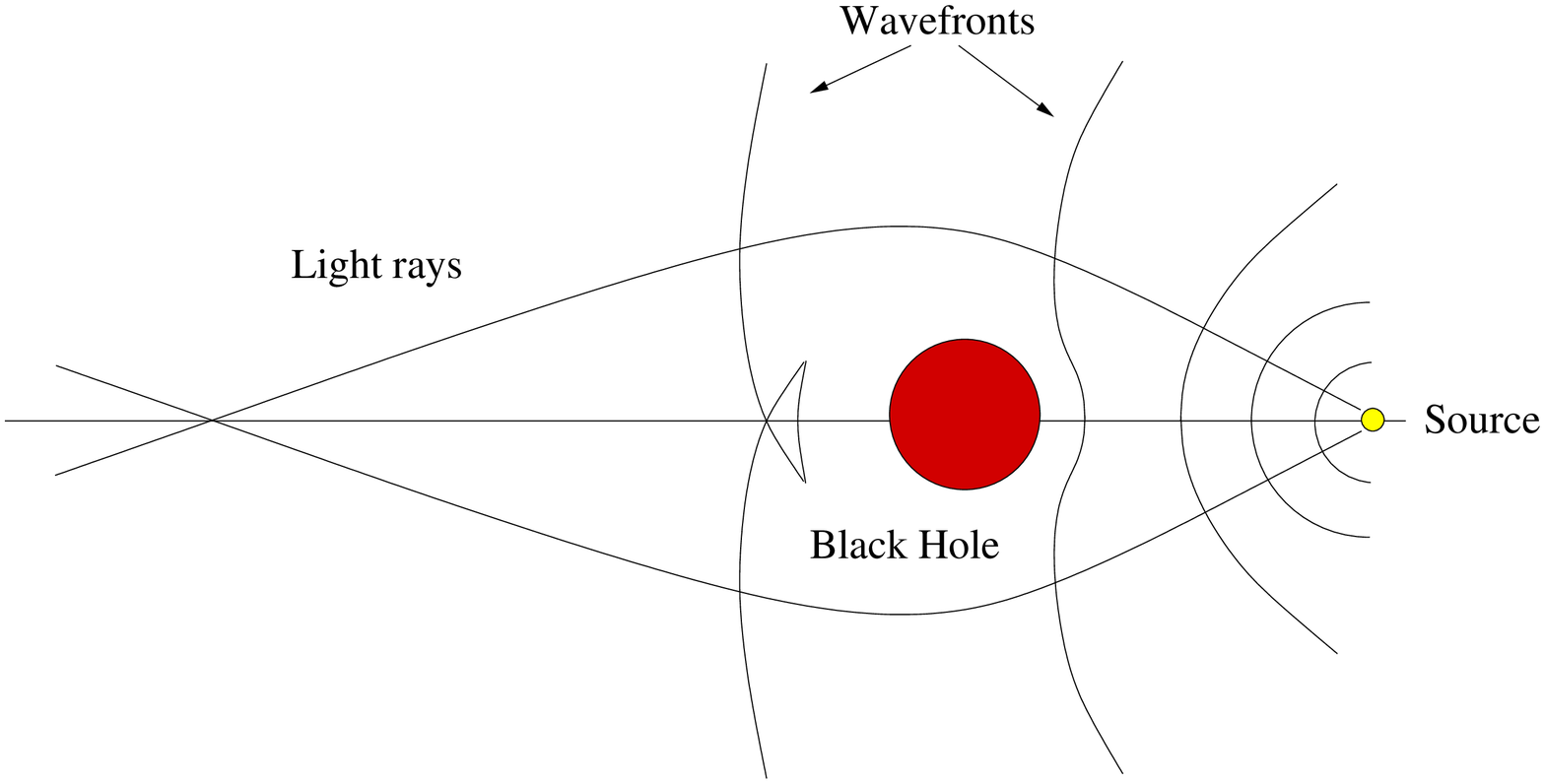}}}
\hskip1cm
\centering{
{\includegraphics[scale=0.5]{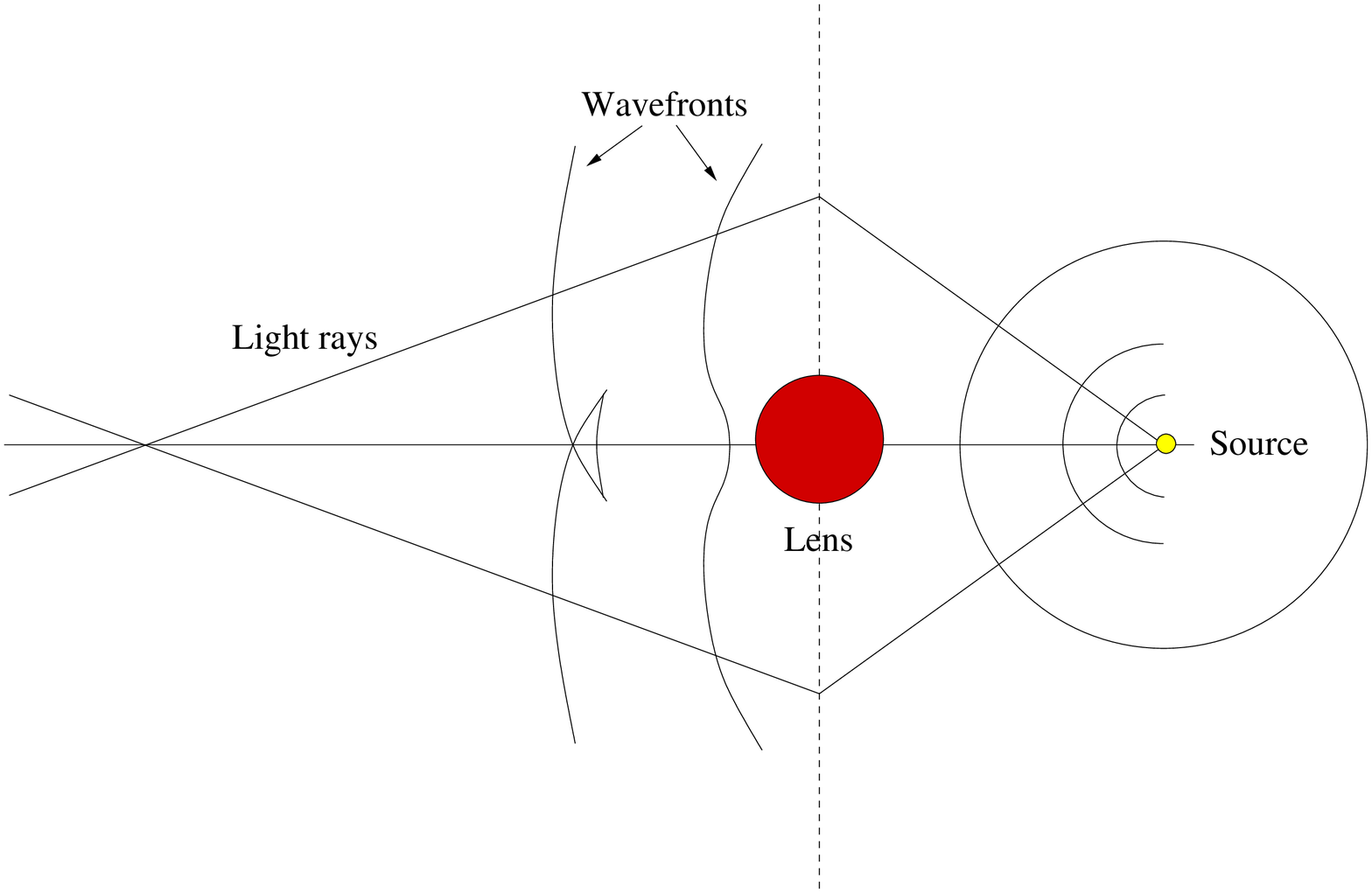}}}
\caption[]{Differences in the evolution of wavefront in presence of a black 
hole (top) and a weak gravitational lens (bottom).}
\label{fig1}
\end{figure*}

\begin{figure*}
\hbox{
\includegraphics[width=6.5cm]{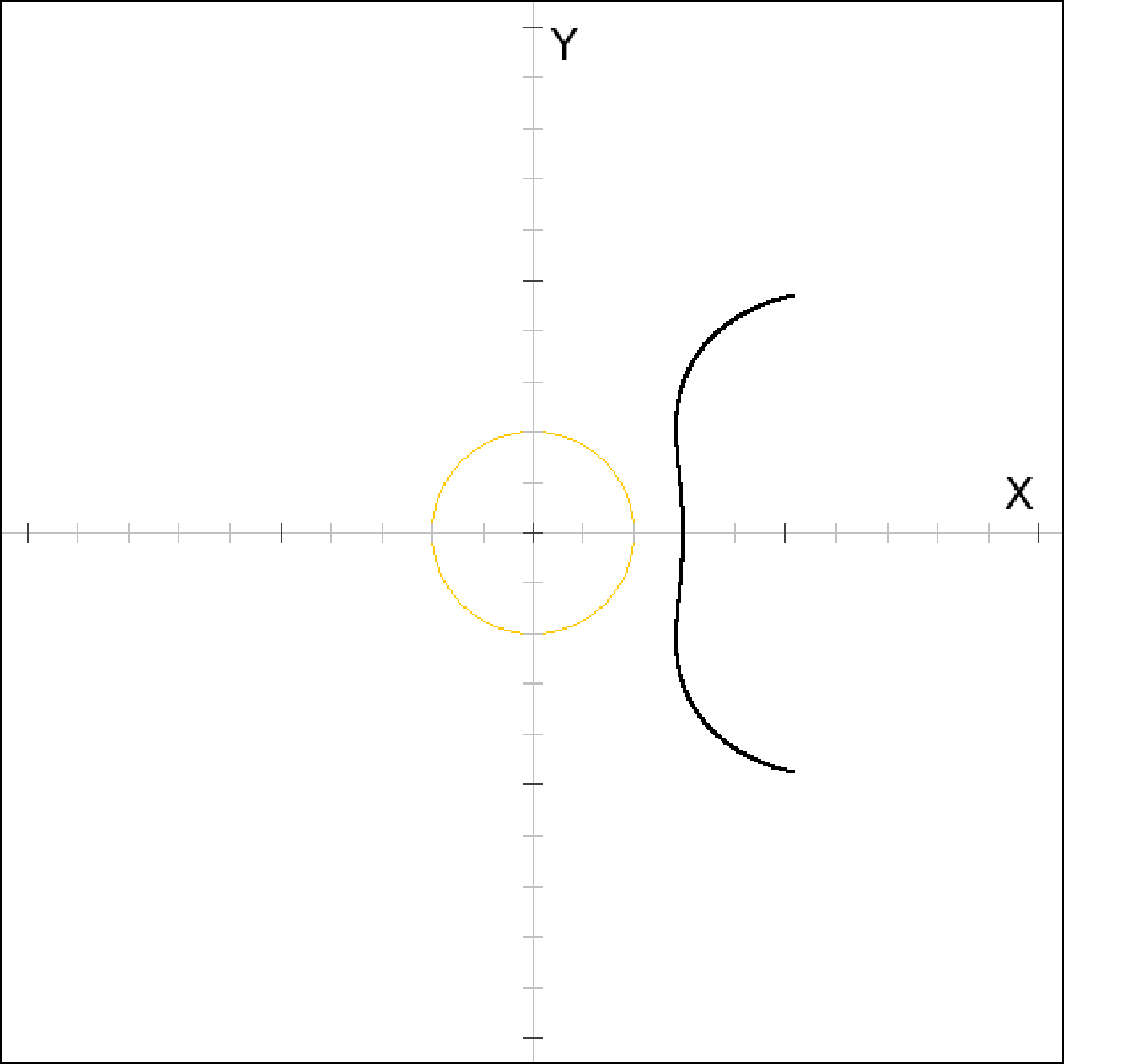} \hspace{5mm}
\includegraphics[width=6.5cm]{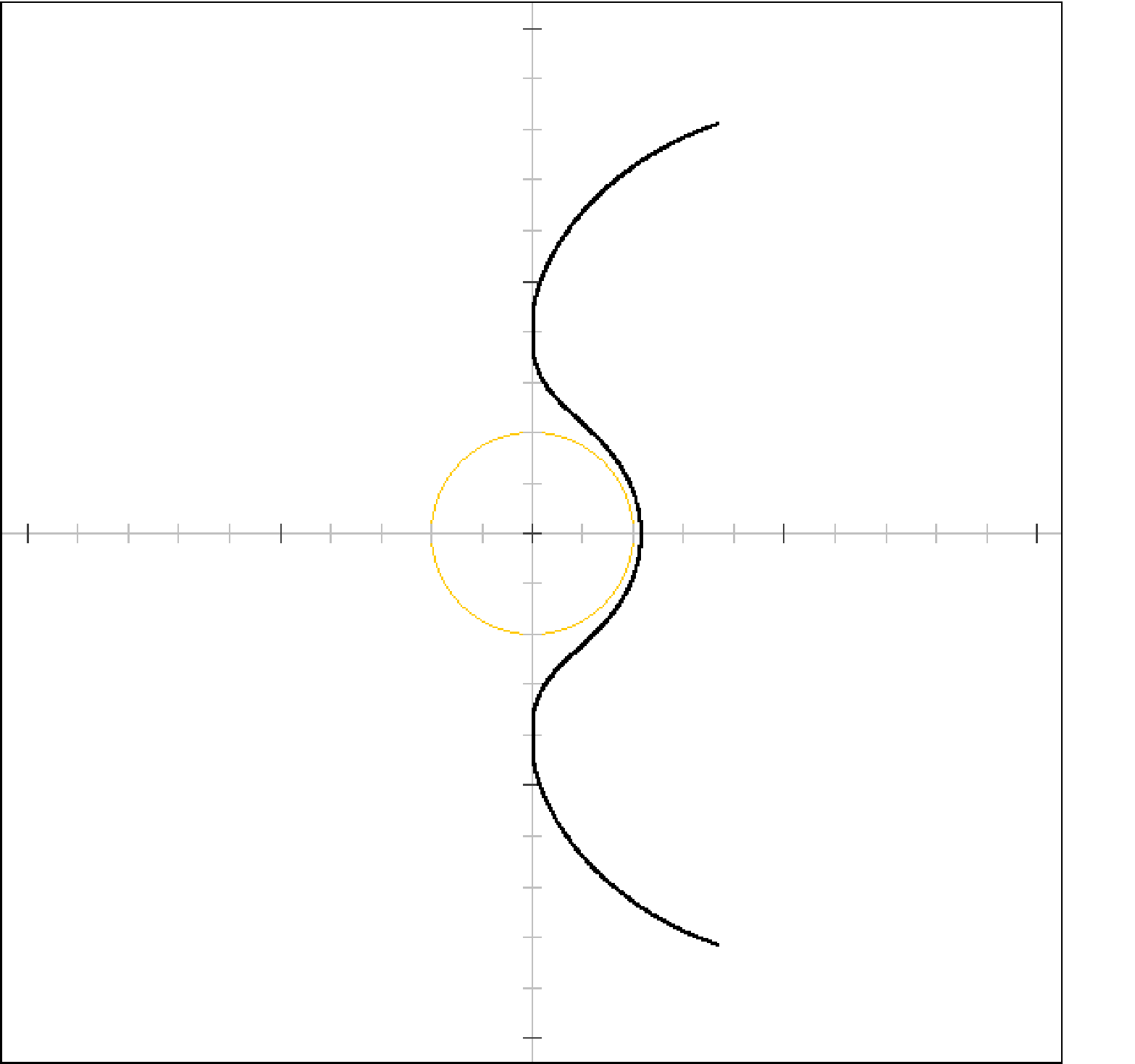}} \vspace{5mm}
\hbox{
\includegraphics[width=6.5cm]{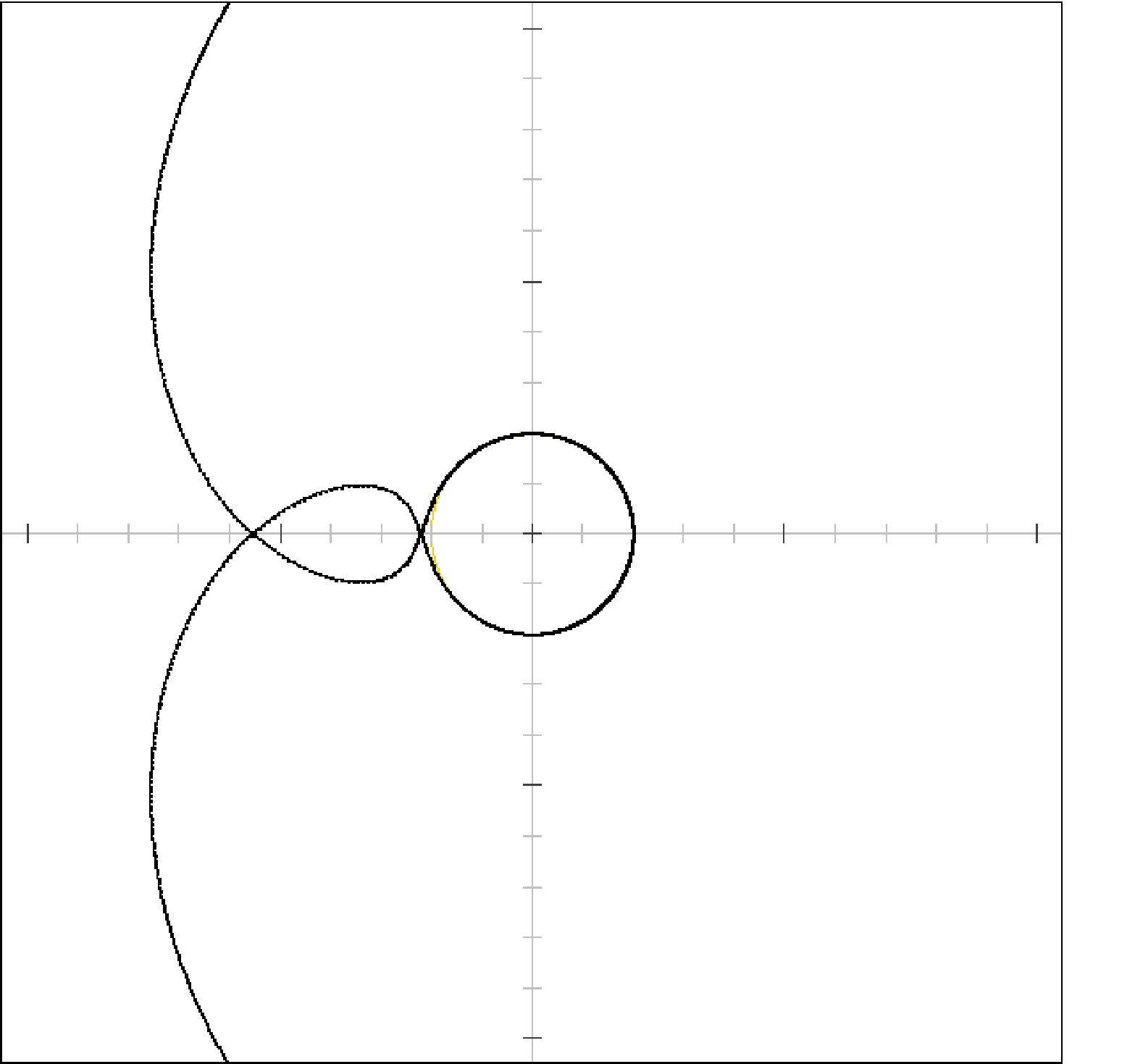} \hspace{5mm}
\includegraphics[width=6.5cm]{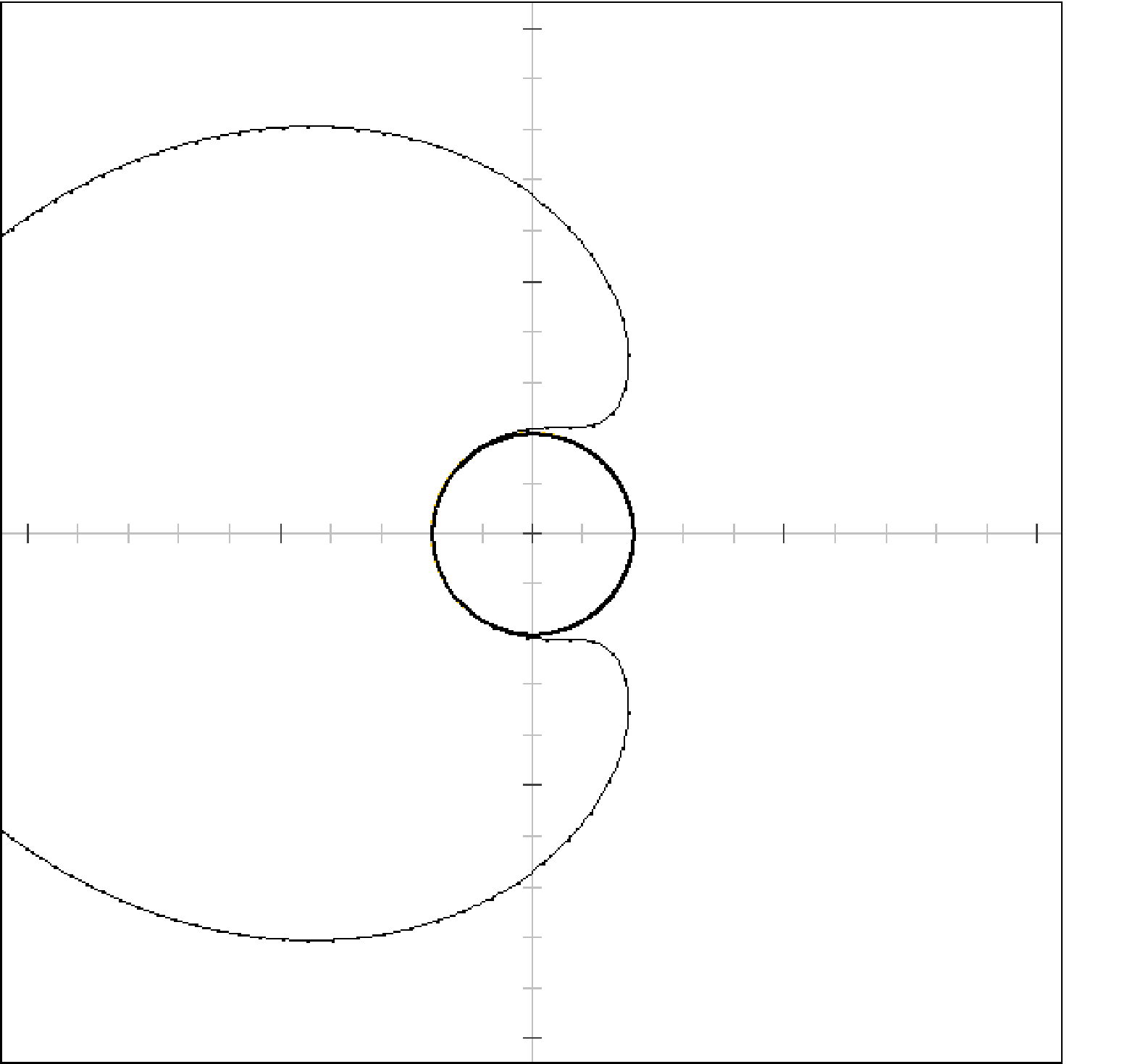}}
\caption[]{Two-dimensional wavefront sequence for the Schwarzschild metric 
($ M = 1 $). The sequence begins on the top left frame. The wavefront is 
moving from the right to the left in the $ x y $ plane. The caustic line 
runs along the $ - x $ axis.}
\label{fig2}
\end{figure*}

\begin{figure*}
\hbox{
\includegraphics[width=6.5cm]{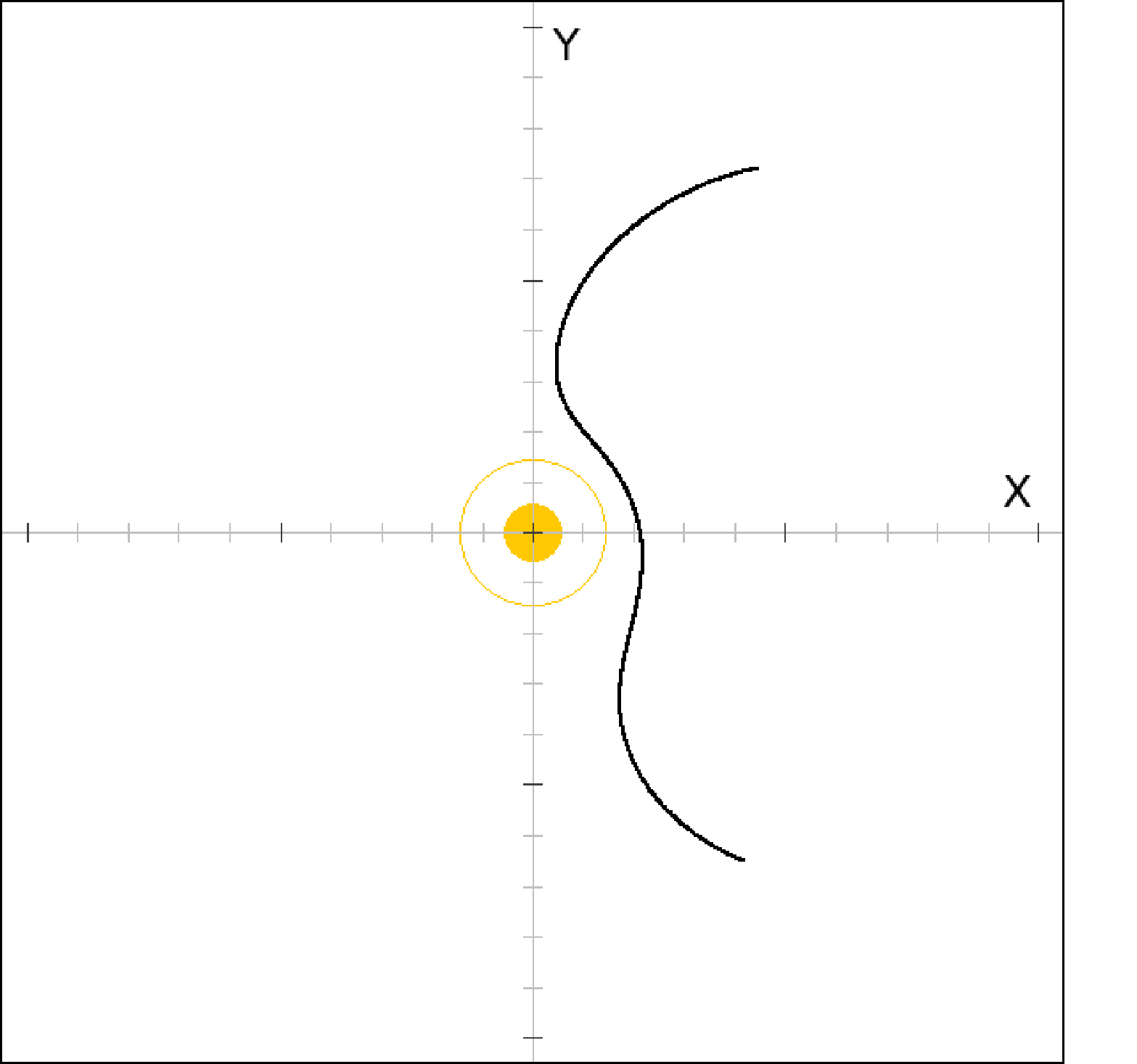} \hspace{5mm}
\includegraphics[width=6.5cm]{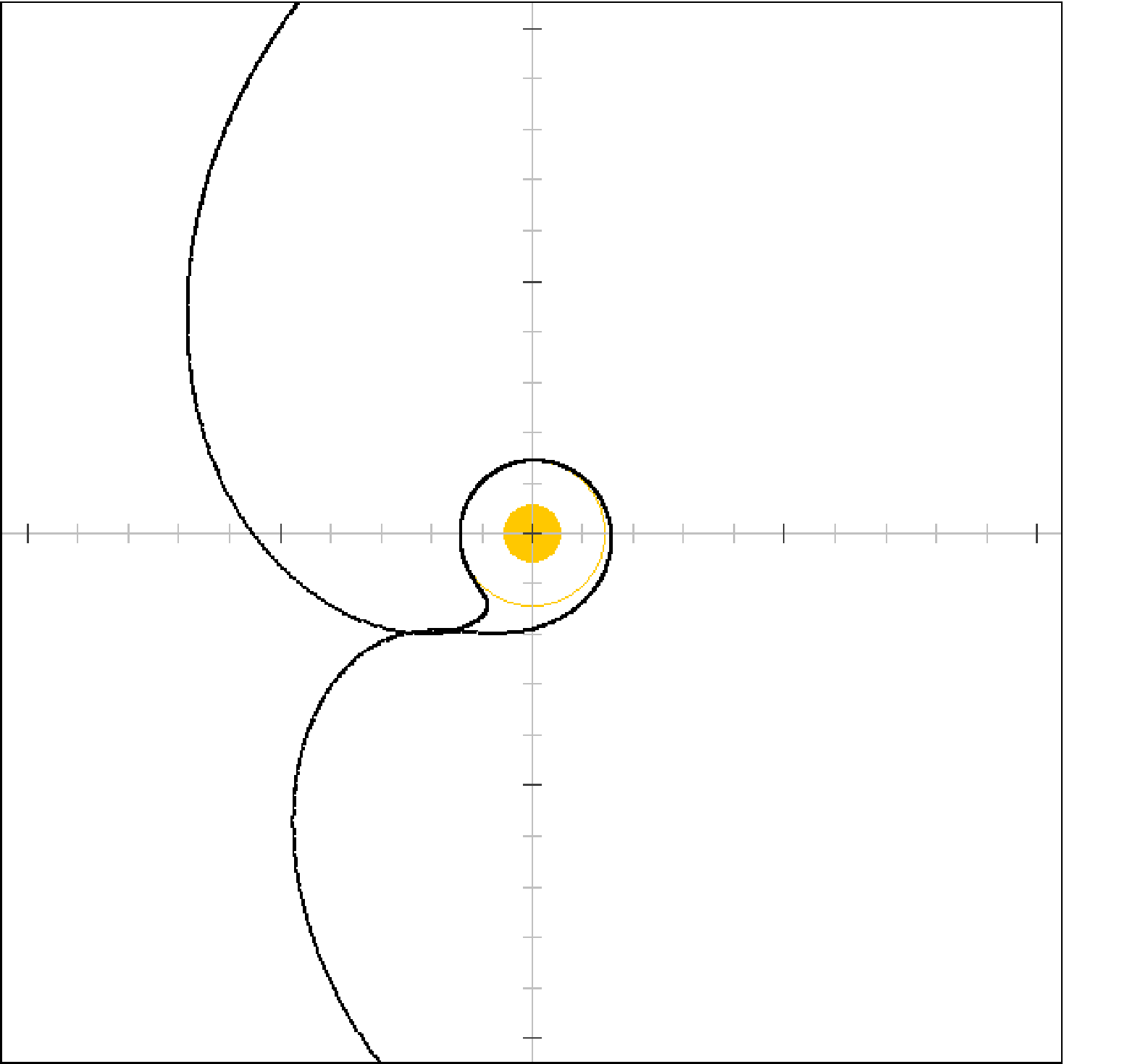}} \vspace{5mm}
\hbox{
\includegraphics[width=6.5cm]{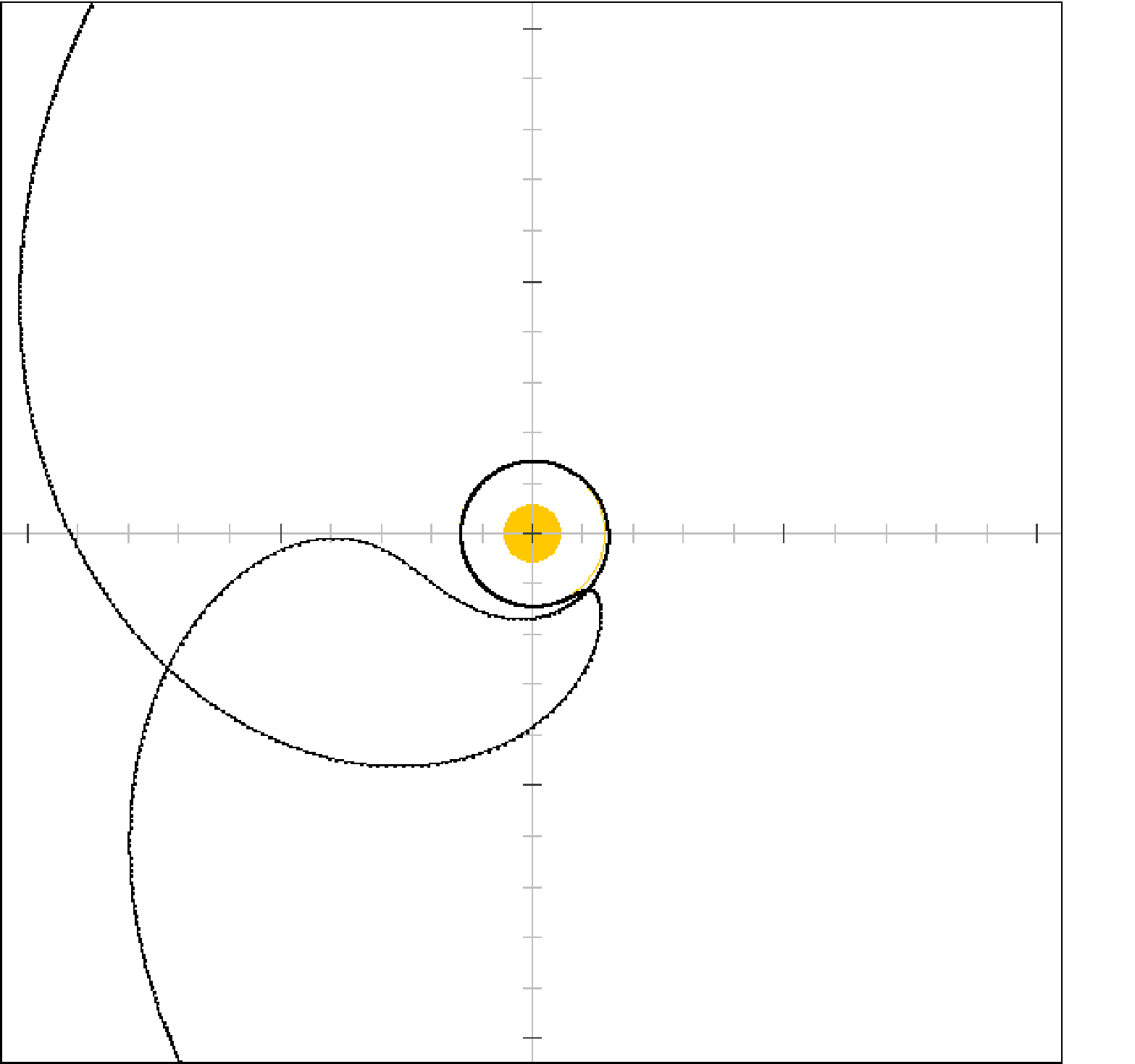} \hspace{5mm}
\includegraphics[width=6.5cm]{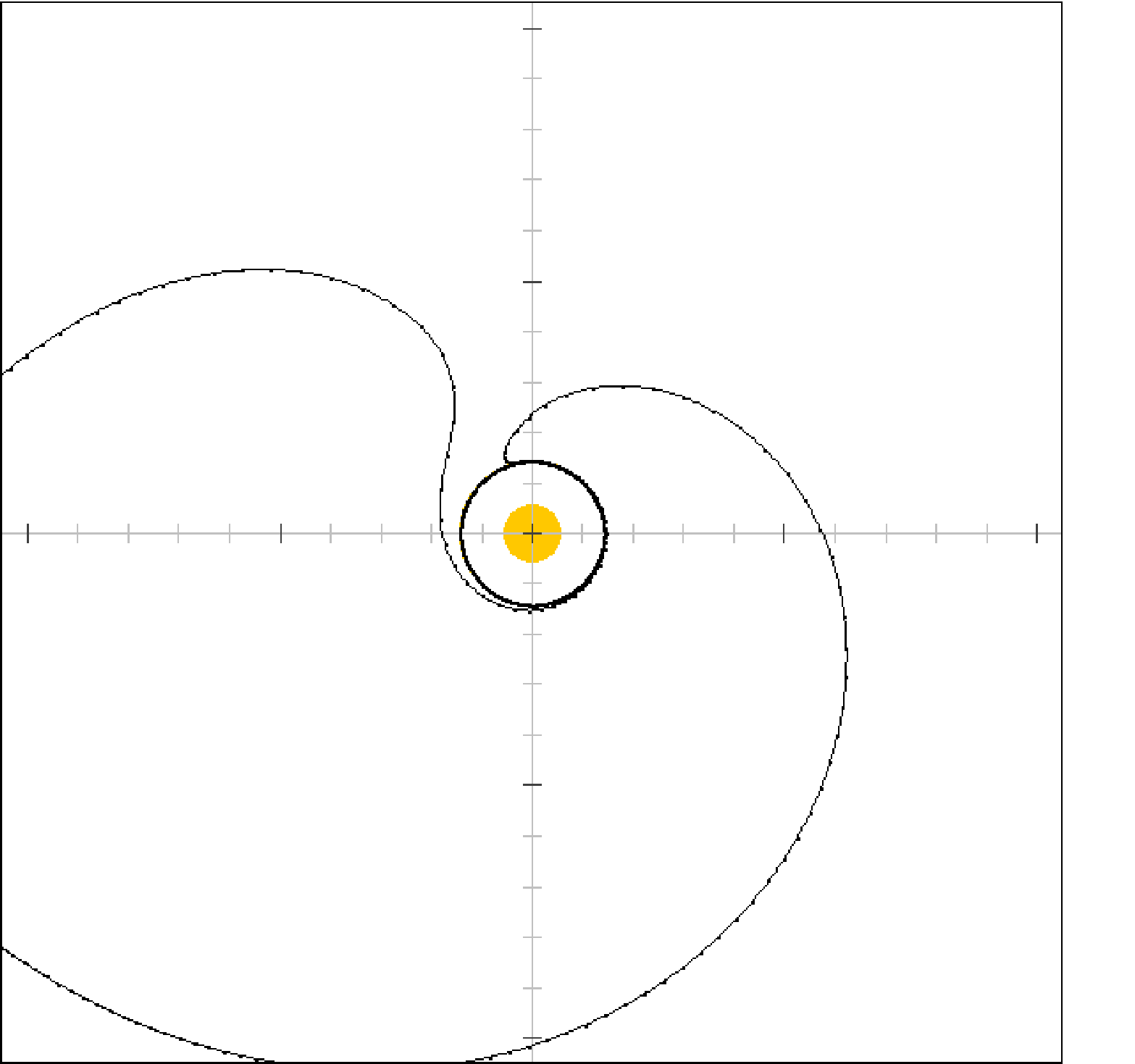}}
\caption[]{Two-dimensional wavefront sequence for the Kerr metric 
($ M = 1 $, $ a = 0.9 $). The sequence begins on the top left frame. 
The wavefront is moving from the right to the left in the $ x y $ plane.
In this case, there is not a caustic line, but we can see caustic points.}
\label{fig3}
\end{figure*}

\begin{figure*}
\hbox{
\includegraphics[width=6.5cm]{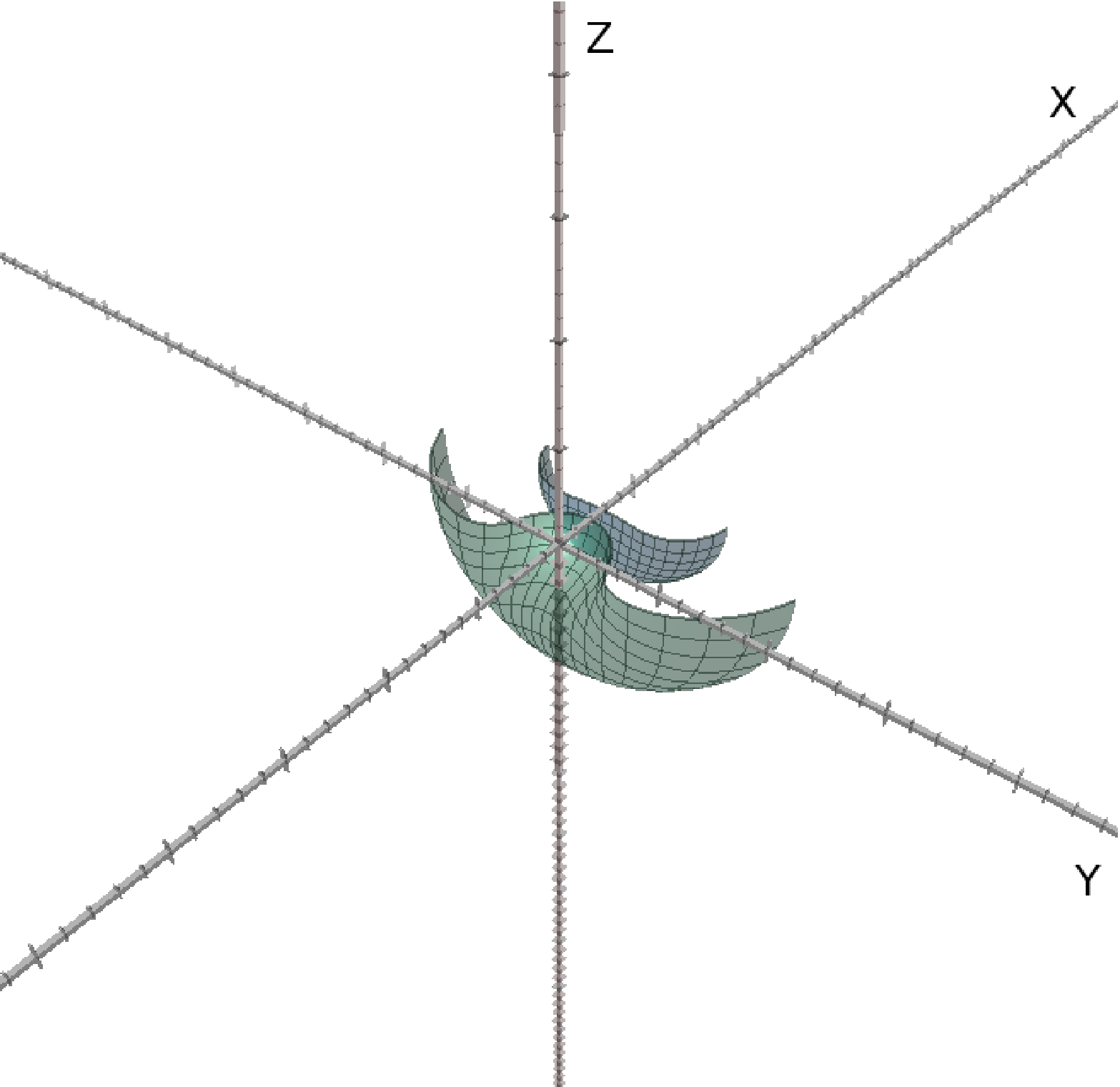} \hspace{5mm}
\includegraphics[width=6.5cm]{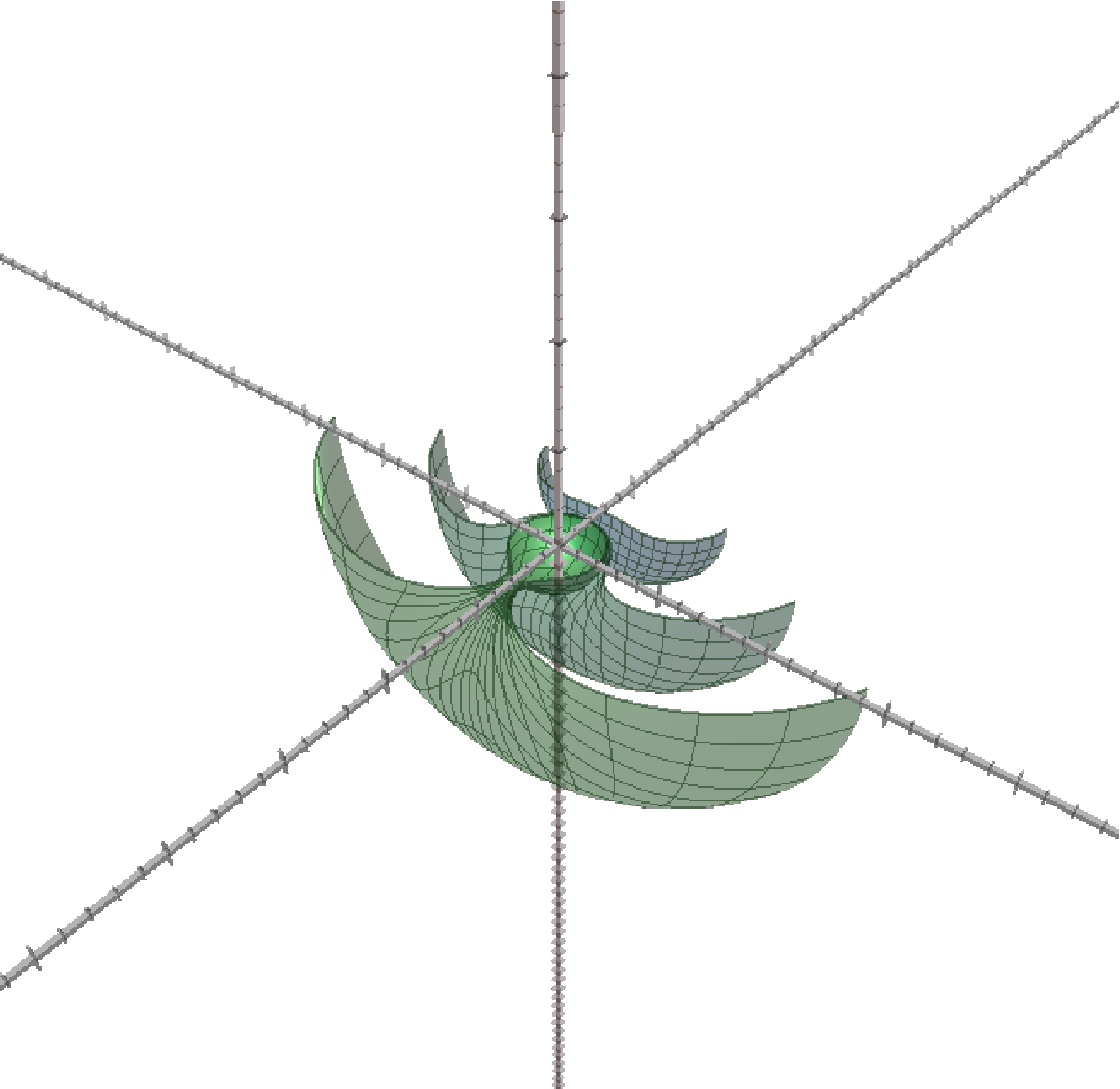}} \vspace{5mm}
\hbox{
\includegraphics[width=6.5cm]{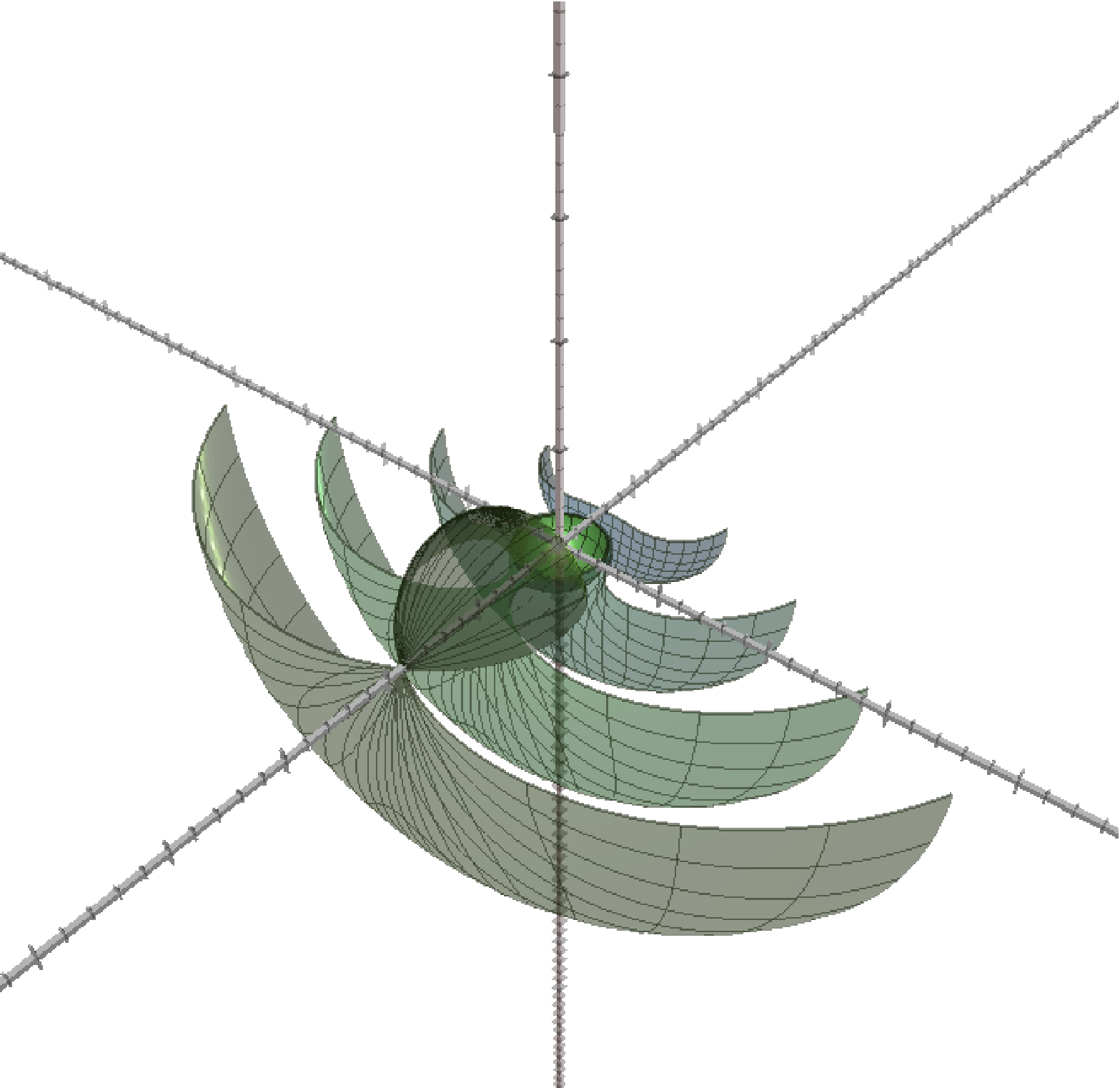} \hspace{5mm}
\includegraphics[width=6.5cm]{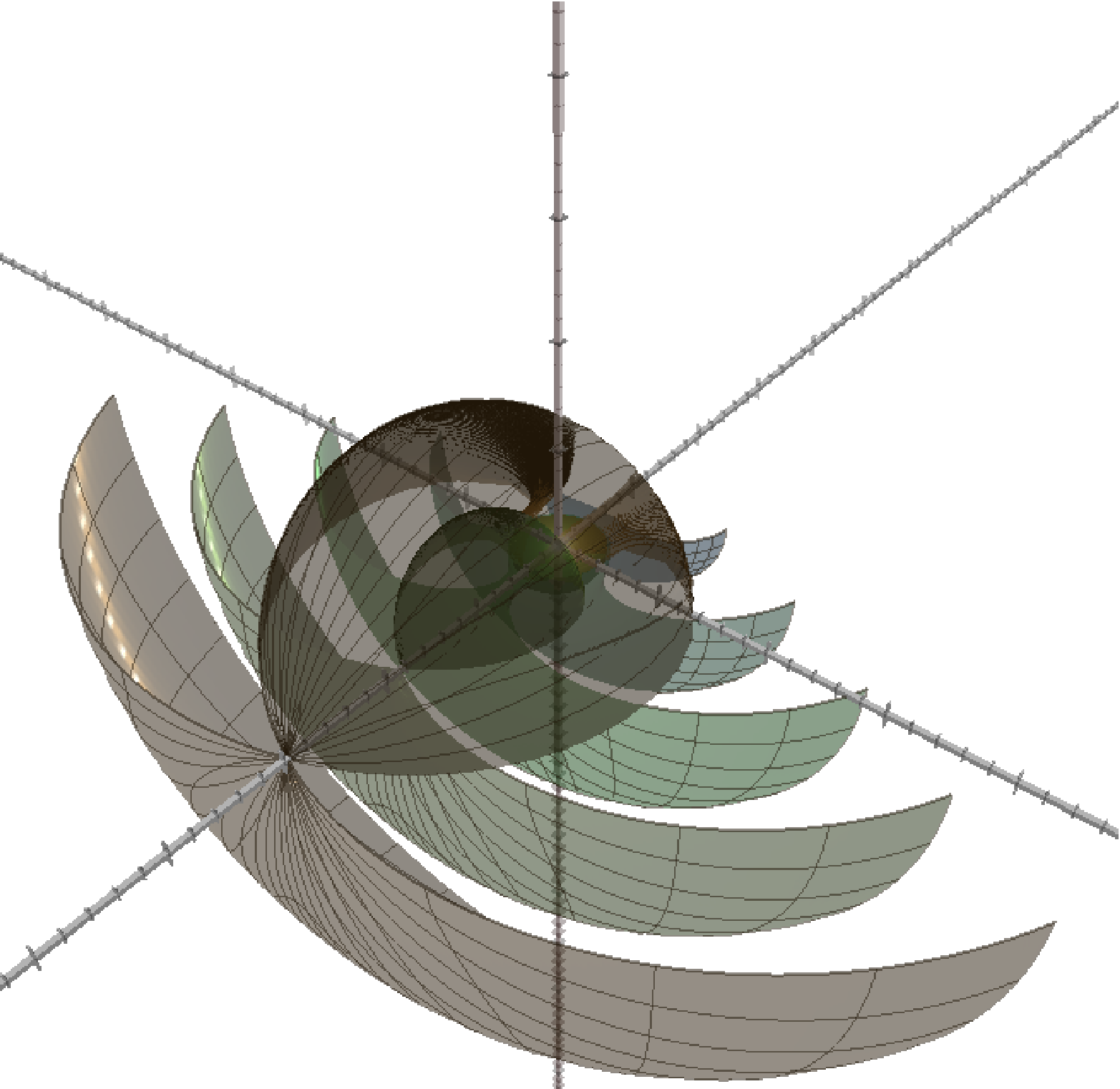}}
\caption[]{Three-dimensional wavefront sequence for the Schwarzschild metric 
($ M = 1 $). The sequence begins on the top left frame.}
\label{fig4}
\end{figure*}

\begin{figure*}
\hbox{
\includegraphics[width=6.5cm]{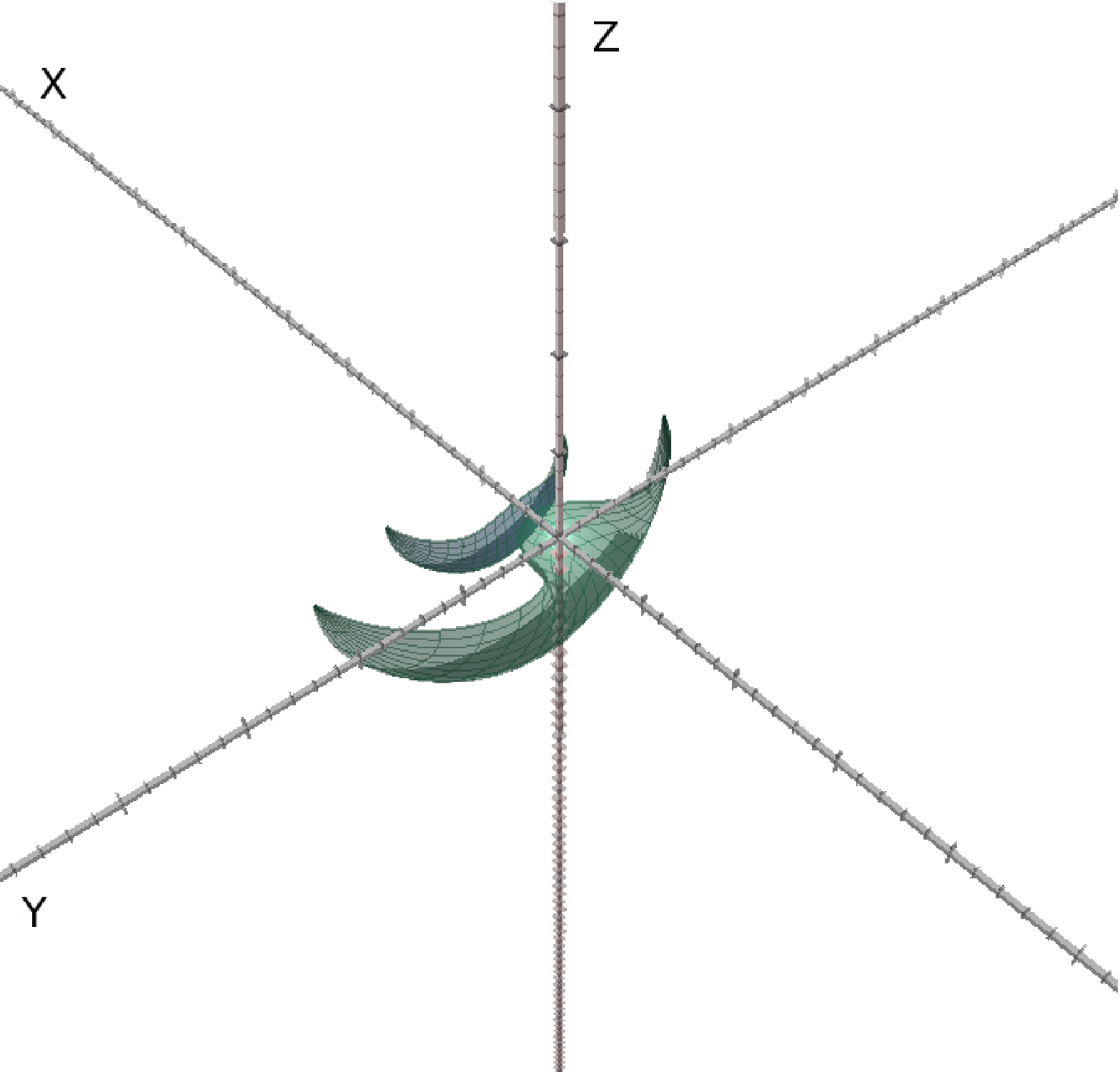} \hspace{5mm}
\includegraphics[width=6.5cm]{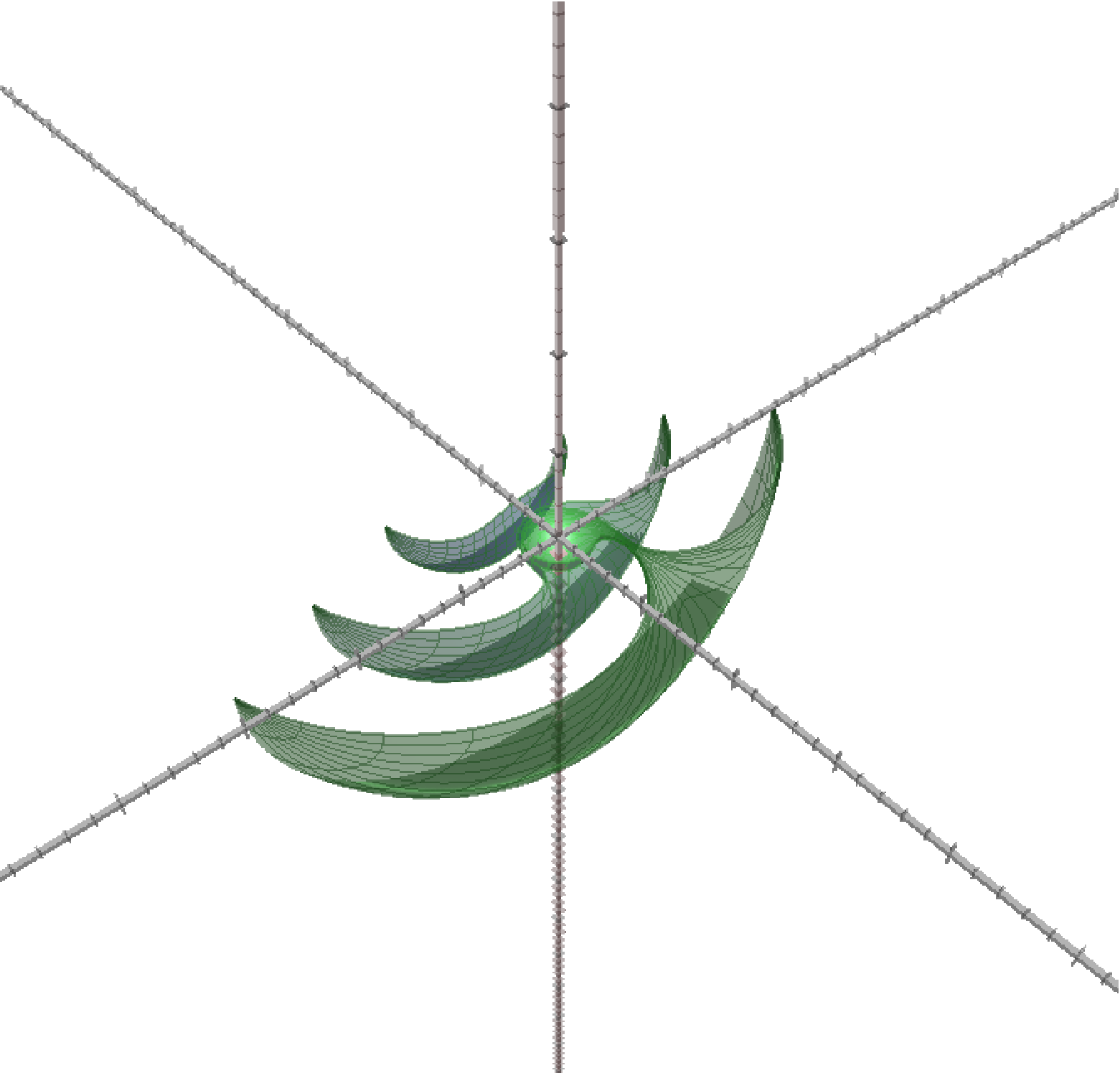}} \vspace{5mm}
\hbox{
\includegraphics[width=6.5cm]{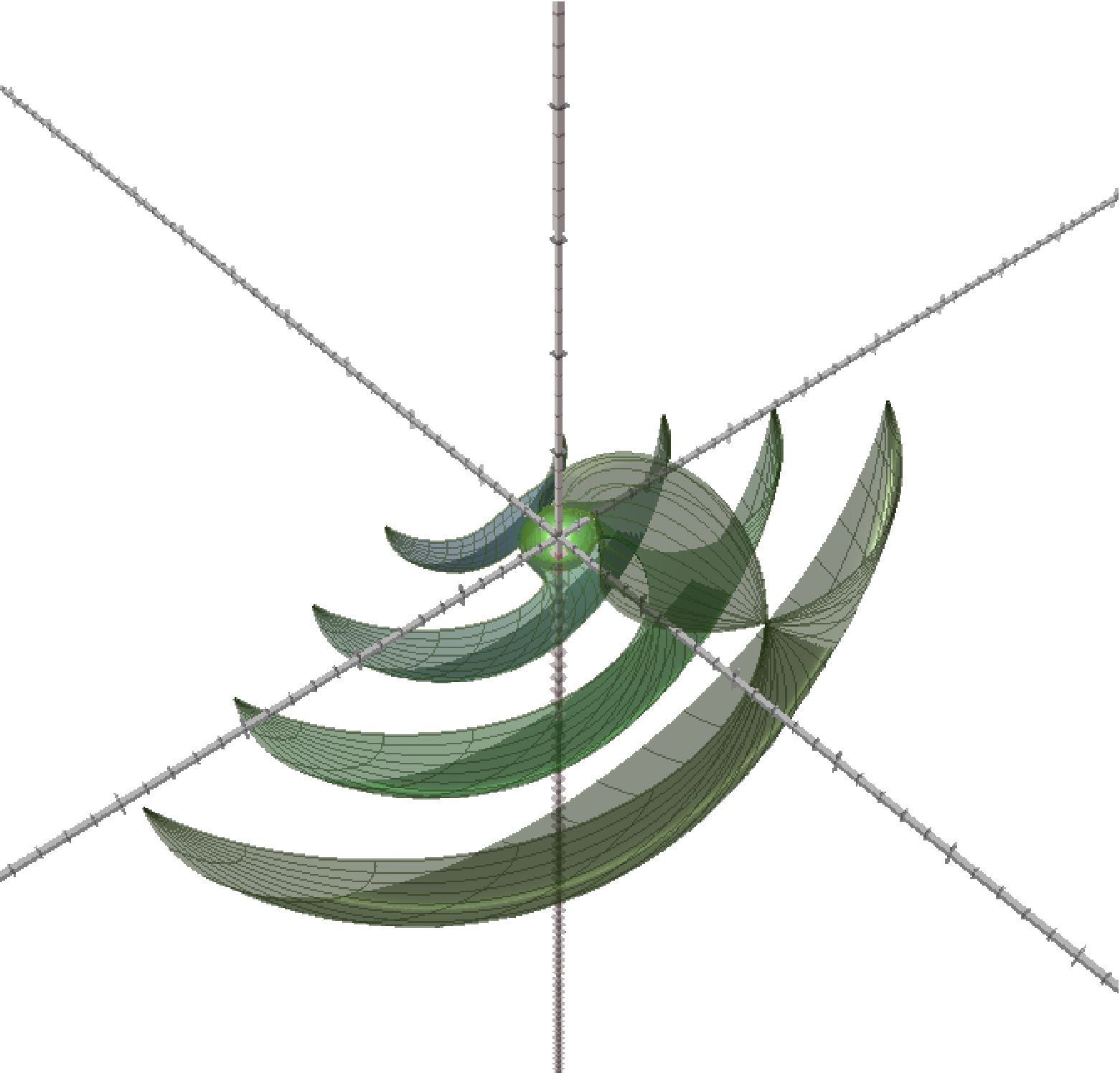} \hspace{5mm}
\includegraphics[width=6.5cm]{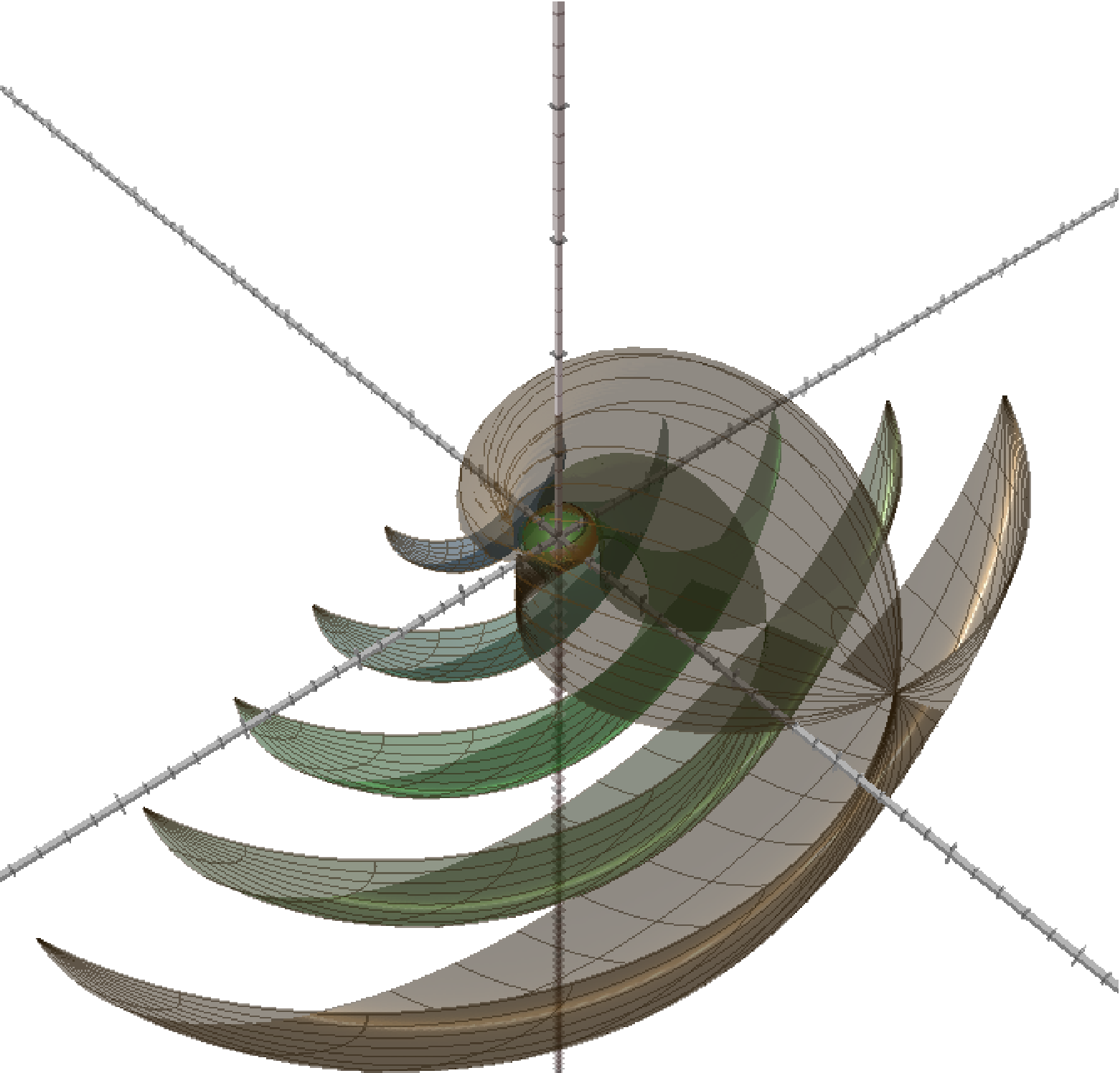}}
\caption[]{Three-dimensional wavefront sequence for the Kerr metric 
($ M = 1 $, $ a = 0.9 $). The sequence begins on the top left frame.}
\label{fig5}
\end{figure*}

\begin{figure*}
\hbox{
\includegraphics[width=6.5cm]{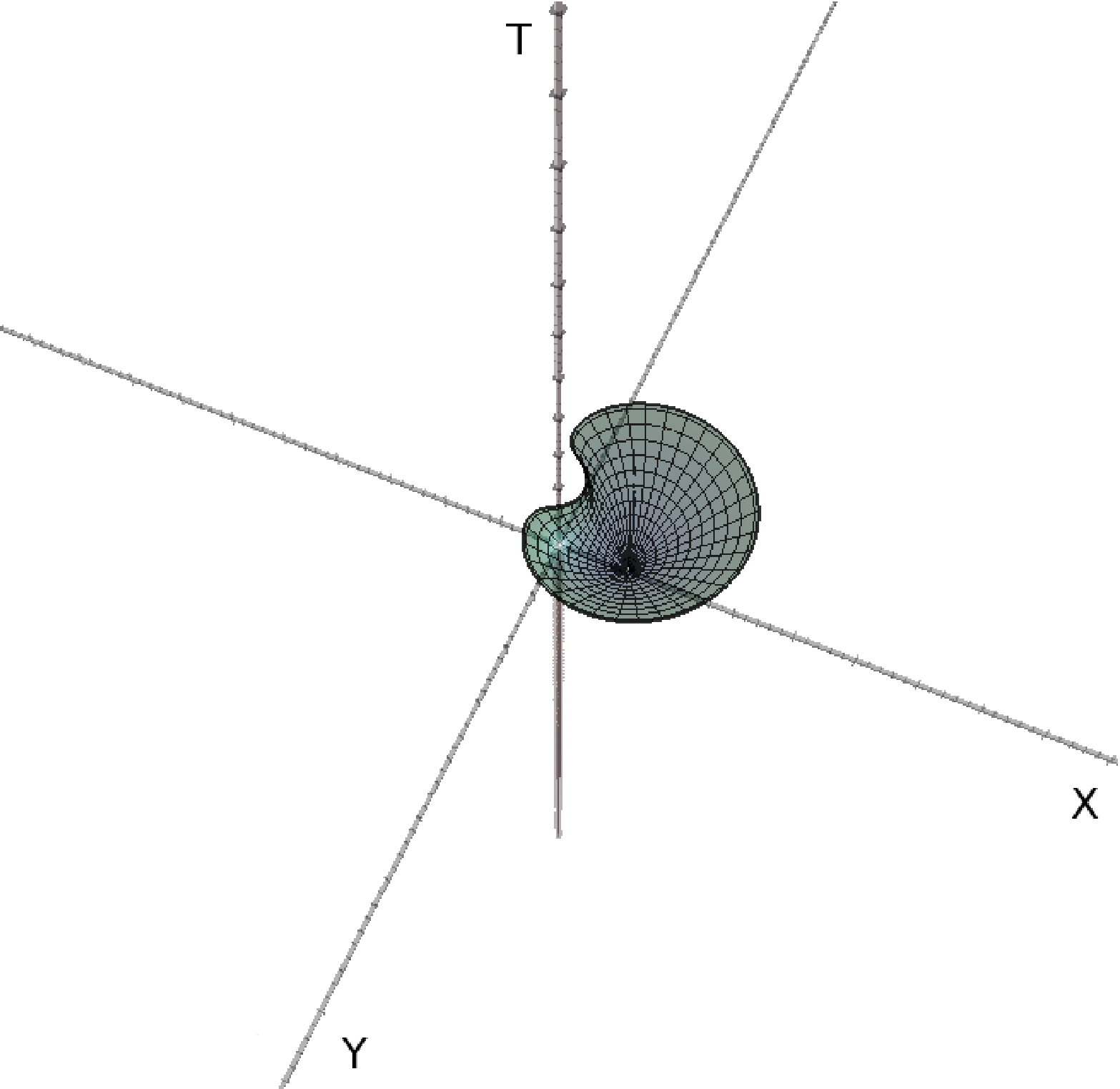} \hspace{5mm}
\includegraphics[width=6.5cm]{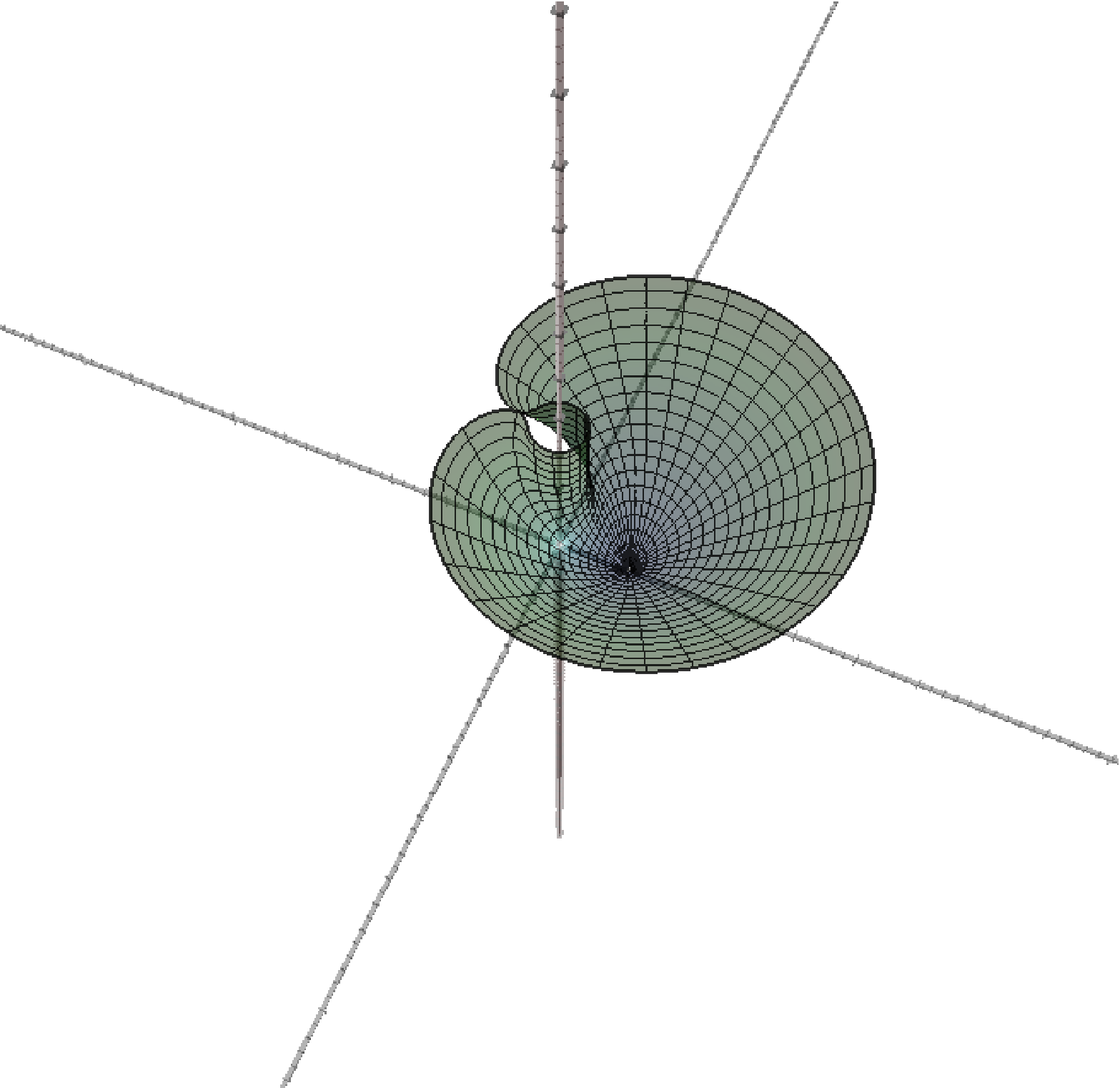}} \vspace{5mm}
\hbox{
\includegraphics[width=6.5cm]{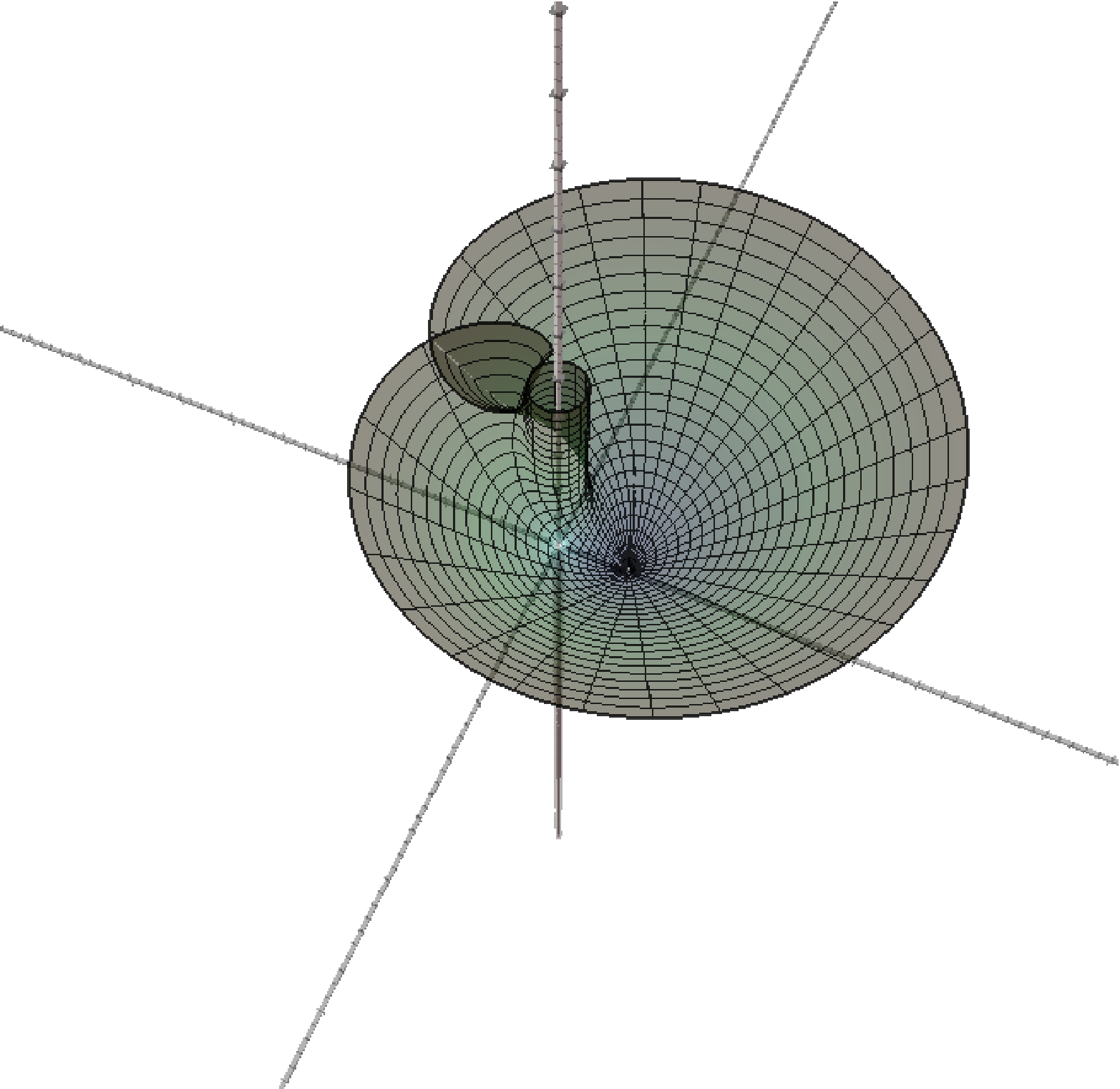} \hspace{5mm}
\includegraphics[width=6.5cm]{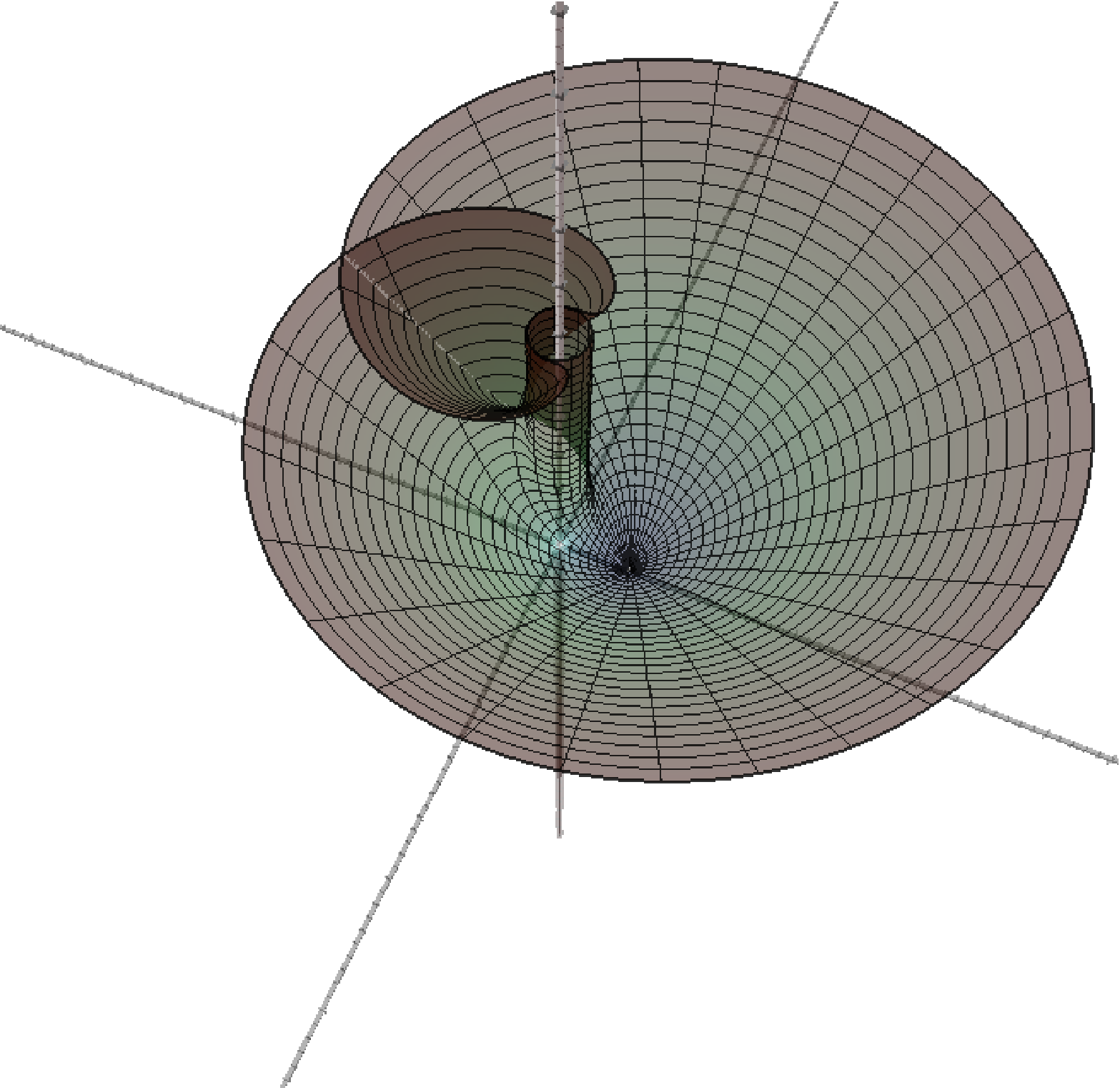}}
\caption[]{Light cone evolution for the Schwarzschild metric 
($ M = 1 $). The sequence begins on the top left frame. The light cone 
evolves from the initial point on the $ x y t $ space.} 
\label{fig6}
\end{figure*}

\begin{figure*}
\hbox{
\includegraphics[width=6.5cm]{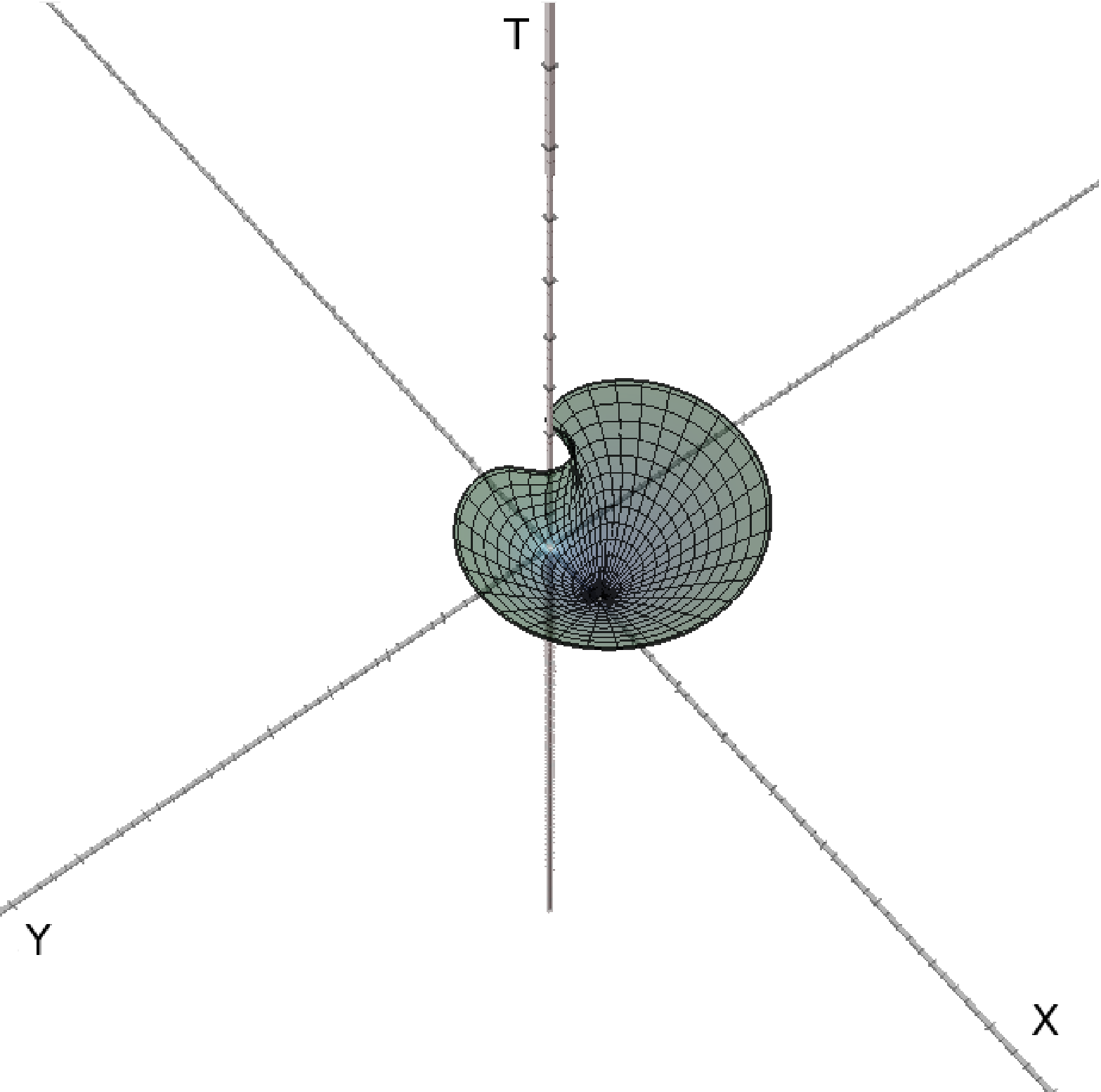} \hspace{5mm}
\includegraphics[width=6.5cm]{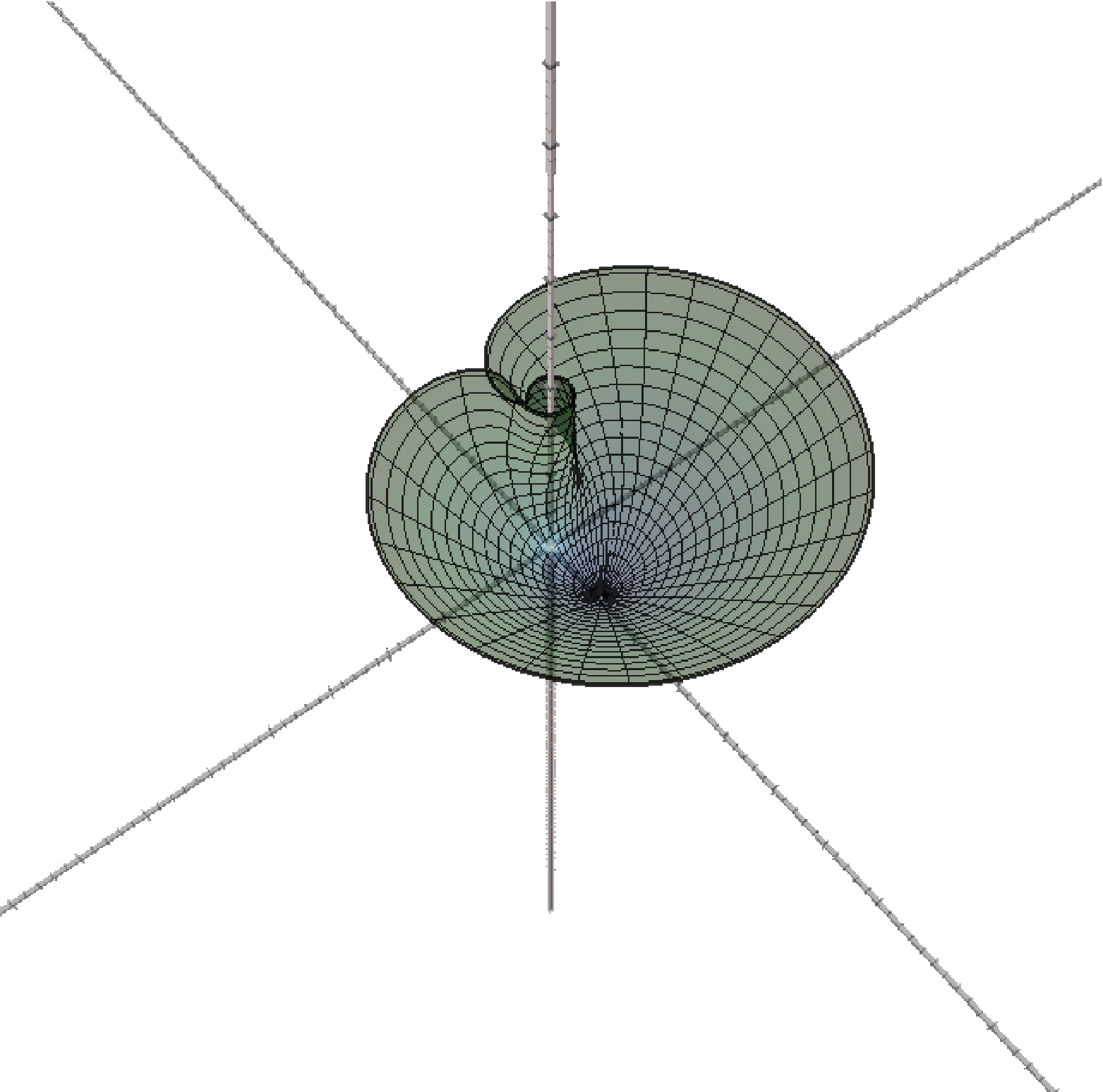}} \vspace{5mm}
\hbox{
\includegraphics[width=6.5cm]{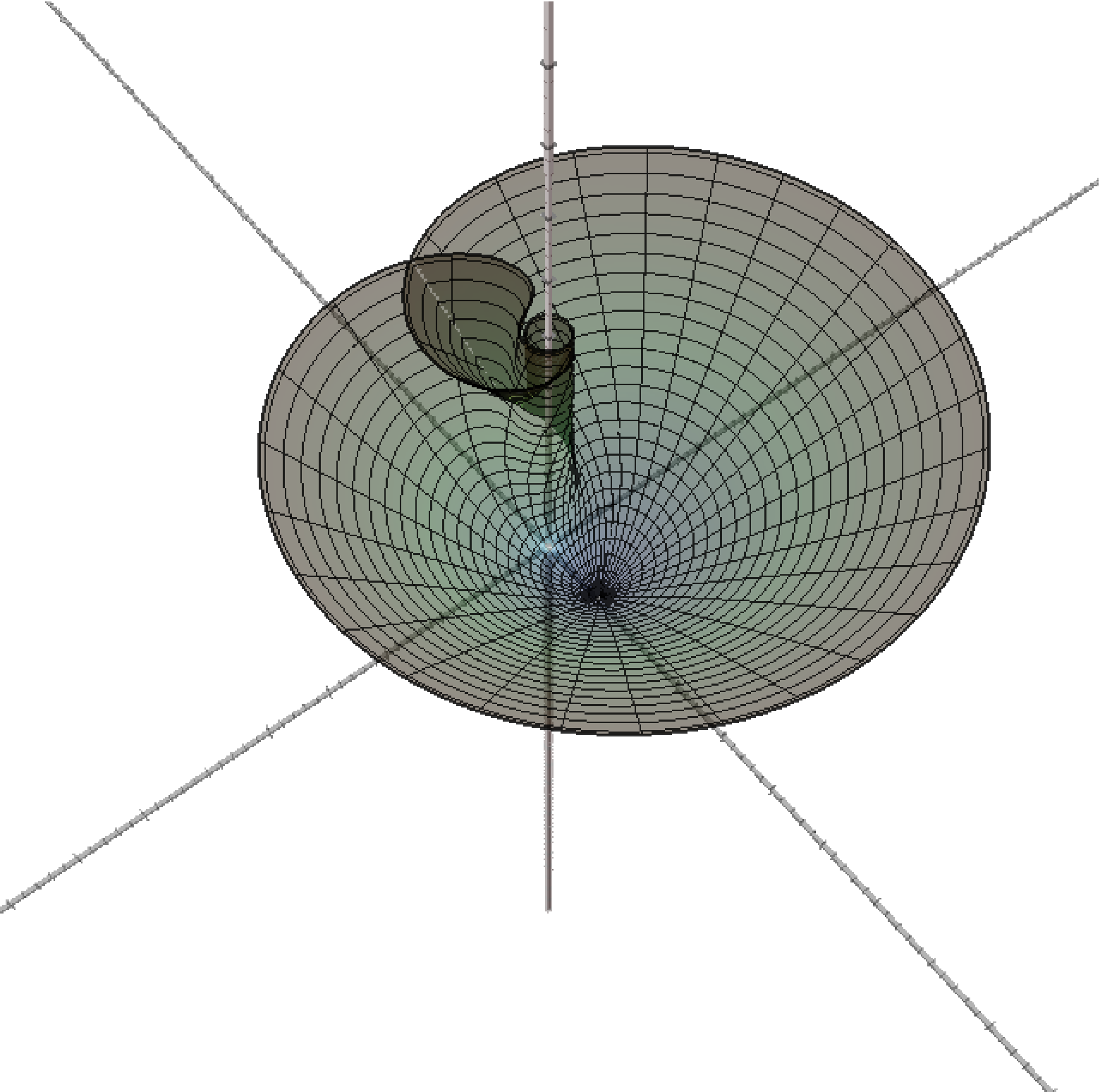} \hspace{5mm}
\includegraphics[width=6.5cm]{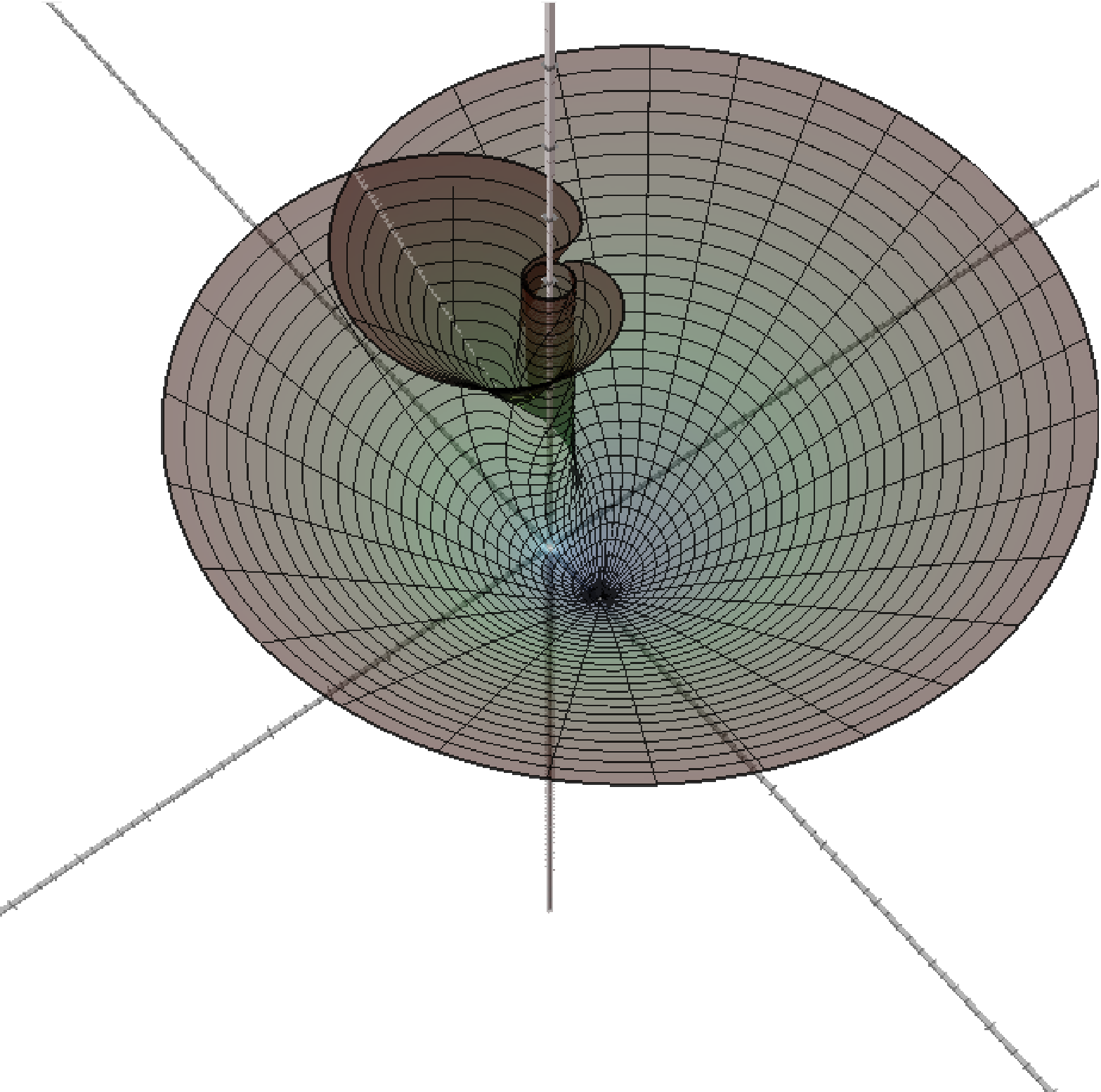}}
\caption[]{Light cone evolution for the Kerr metric 
($ M = 1 $, $ a = 0.9 $). The sequence begins on the top left frame. 
The light cone evolves from the initial point on the $ x y t $ space.}
\label{fig7}
\end{figure*}

\end{document}